\documentclass[%
 aip,
 amsmath,amssymb,
 reprint,% For two-column format
 author-year,%
]{revtex4-2}

\usepackage{graphicx}% Include figure files
\usepackage{dcolumn}% Align table columns on decimal point
\usepackage{bm}% bold math

\usepackage[utf8]{inputenc}
\usepackage[T1]{fontenc}
\usepackage{mathptmx}
\usepackage{etoolbox}
\usepackage{tikz-cd}
\usepackage{siunitx}
\usepackage{calc}
\usepackage[hidelinks]{hyperref}
\usepackage{afterpage}

\newcommand{\Rey}{\mbox{\textit{Re}}}
\newcommand{\mean}[3]{\left\langle\left\{#1\right\}_{#2}^#3\right\rangle}
\newcommand{\mat}[1]{\mathbf{\mathsf{#1}}} % Matrix
\DeclareMathOperator*{\argmax}{arg\,max}
\newcommand{\pluscross}{\ensuremath{\times}\kern-0.78em{\ensuremath{+}}}

\newcommand{\renum}{390} % Reynolds number
\newcounter{nx}
\newcounter{ny}
\newcounter{nz}
\newcounter{nsamples}
\newcounter{nplane}
\newcounter{ngrid}
\newcounter{nepoddom}
\newcounter{ndpoddom}
\newcounter{allsamples}
\makeatletter
\def\@email#1#2{%
 \endgroup
 \patchcmd{\titleblock@produce}
  {\frontmatter@RRAPformat}
  {\frontmatter@RRAPformat{\produce@RRAP{*#1\href{mailto:#2}{#2}}}\frontmatter@RRAPformat}
  {}{}
}%
\makeatother
\begin{document}

\preprint{AIP/123-QED}

\title{Dissipation-optimized Proper Orthogonal Decomposition}
\author{P. J. Olesen}
\email{pjool@dtu.dk}
\affiliation{Department of Civil and Mechanical Engineering, Technical University of Denmark, 2800 Kgs.\ Lyngby, Denmark}%
\author{A. Hod\v{z}i\'c}%
\affiliation{Department of Civil and Mechanical Engineering, Technical University of Denmark, 2800 Kgs.\ Lyngby, Denmark}
\author{S. J. Andersen}
\affiliation{Department of Wind and Energy Systems, Technical University of Denmark, 2800 Kgs.\ Lyngby, Denmark}
\author{N. N. Sørensen}
\affiliation{Department of Wind and Energy Systems, Technical University of Denmark, 4000 Roskilde, Denmark}
\author{C. M. Velte}
\affiliation{Department of Civil and Mechanical Engineering, Technical University of Denmark, 2800 Kgs.\ Lyngby, Denmark}

\date{\today}

\begin{abstract}
We present a formalism for dissipation-optimized 
decomposition of the strain rate tensor (SRT) of turbulent flow data using Proper Orthogonal Decomposition (POD). The formalism includes a novel \textit{inverse} spectral SRT operator allowing the mapping of the resulting SRT modes to corresponding velocity fields, which enables a complete dissipation-optimized reconstruction of the velocity field. Flow data snapshots are obtained from a direct numerical simulation of a turbulent channel flow with friction Reynolds number $\Rey_{\tau}=\renum$. The lowest dissipation-optimized POD (d-POD) modes are compared to the lowest conventional turbulent kinetic energy (TKE) optimized POD (e-POD) modes. The lowest d-POD modes show a richer small-scale structure, along with traces of the large-scale structure characteristic of e-POD modes, indicating that the former capture structures across a wider range of spatial scales. Profiles of both TKE and dissipation are reconstructed using both decompositions, and reconstruction convergences are compared in all cases. Both TKE and dissipation are reconstructed more efficiently in the dissipation-rich near-wall region using d-POD modes, and in the TKE-rich bulk using e-POD modes. Lower modes of either decomposition tend to contribute more to either reconstructed quantity. Separating each term into eigenvalues and factors relating to the inherent structures in each mode reveals that higher e-POD modes tend to encode more dissipative structures, whereas the structures encoded by d-POD modes have roughly constant inherent TKE content, supporting the hypothesis that structures encoded by d-POD modes tend to span a wide range of spatial scales.
\end{abstract}

\maketitle

\section{Introduction}
Viscous dissipation is a parameter crucial for understanding and modeling dynamics of turbulent flows \citep{pope2000turbulent}. Existing approaches to flow decompositions used e.g.\ for gaining insight into flow dynamics and for formulating reduced order models (ROMs) typically educe structures based on their turbulent kinetic energy (TKE) content, neglecting dissipative structures. Various techniques have been developed to compensate for unresolved dissipative structures in ROMs; however, explicitly optimizing the flow decomposition with respect to dissipation would enable closer studies of dissipative structures, while also providing an avenue for constructing more robust ROMs.

The application of proper orthogonal decomposition (POD) to turbulent flows was pioneered by \citet{Lumley1967} with the aim of objectively identifying and characterizing dominant large scale flow structures. Applying this classical POD to a velocity fluctuation field ensemble produces an orthogonal basis spanning the ensemble such that reconstructing the fluctuation field using a given number of modes reproduces the optimal amount of mean TKE compared to any other decomposition. In this sense, a POD-based modal expansion is optimally robust against truncation \citep[see][]{berkooz1993proper,holmes2012turbulence}. This explicit optimization with respect to TKE is what leads to large-scale structures being prioritized in the decomposition.

\citet{Gatski1992} applied the classical POD to channel flow data obtained from a Direct Numerical Simulation (DNS) and reconstructed the TKE, the shear stresses, and the dissipation, demonstrating that while the former two could be reconstructed accurately using relatively few modes, the reconstruction of the dissipation required a larger number of modes. This was taken as a confirmation of the work of \citet{ukeiley1992multifractal}, which suggested that the ordering of POD modes by decreasing mean TKE contribution was equivalent to ordering by decreasing characteristic length scales in the flow structures described by the modes. Dissipation, being related to small scale motions, would therefore be poorly reconstructed using an expansion favoring large scales. The parallel between POD modes and length scales was extended by \citet{couplet2003intermodal}, who showed that for a flow past a backward-facing step, the concept of the energy cascade from larger to smaller length scales described by \citet{richardson1922weather} also applied to POD modes in this flow; their analysis of triadic terms in the Galerkin-projected energy transport equation showed a local net flow of TKE from lower towards higher POD modes. The implication, in agreement with the findings of \citet{Gatski1992}, is that higher POD modes are needed to describe the small-scale velocity fluctuations.

\citet{ali2016focused} applied multifractal analysis to wind turbine wake dissipation signals reconstructed using either the lowest several POD modes, associated with larger length scales, or the remaining modes, associated with smaller length scales. It was demonstrated that the multifractal structure of the full dissipation signal was reproduced more accurately using higher modes than using lower modes, supporting the notion that dissipation information is encoded in higher modes. On the other hand, it was shown by \citet{Lee2020} that for an anisotopic 2D flow, reordering POD modes by their mean velocity gradient norm contribution rather than their mean TKE contribution resulted in only a small degree of rearrangement. This was interpreted as evidence against modal scale ordering and modal cascade, and it was suggested instead that each POD mode captures a wide range of dynamical scales; however, this result must be viewed in the light of the different dynamics in play for 2D turbulence compared to the 3D case.

The association of small-scale structures with higher and less energetic POD modes suggests that modal optimality with respect to TKE causes small-scale fluctuations to be underrepresented: a representation that prioritizes TKE-rich larger scales will be inefficient at reconstructing the smaller scales characterizing dissipation. This trade-off means that POD-based ROMs lack accurate representation of small-scale fluctuations important in determining dissipation, which is a central parameter in turbulence theory and modeling. This under-representation of small scales has often been associated with model instability \citep[see][]{bergmann2009enablers}, although this claim has been challenged by \citet{grimberg2020stability} who demonstrated that instabilities are inherent to the Galerkin formalism rather than a consequence of the scales represented in the basis. This notwithstanding, model accuracy relies on the choice of basis: since turbulent dynamics are characterized by interactions between a wide range of scales, even low energetic structures generally play an important role in determining the dynamics of the flow. 

To achieve accurate POD-based ROMs one may therefore include corrections to account for the neglected modes. \citet{bergmann2009enablers} presented an overview of approaches to address this issue, of which only a few will briefly be recounted here. In the earliest work on POD-based ROMs, \citet{aubry1988dynamics} used a generalized Heisenberg model in which the effective viscosity was adjusted to correct for unresolved modes. \citet{iollo2000two} described two avenues to stabilize ROMs, namely compensation for unresolved modes by the addition of an explicit dissipation term, and inclusion of the velocity gradient in the norm with respect to which optimization was performed, though the latter approach was not realized in that work. Furthermore, \citet{Lee2020} formulated a 2D ROM in terms of two POD bases, supplementing the TKE-optimized velocity fluctuation field representation with an enstrophy-optimized gradient field representation. \citet{PS-ROM2022} showed how stability can be achieved by ensuring that the non-linear interactions across modes are correctly maintained when utilizing a global POD basis.

In the present work we build upon the concept presented by \citet{Lee2020}, introducing an explicitly dissipation-optimizing POD formulation (d-POD). The resulting d-POD modes span the set of strain rate tensors (SRTs) derived from the ensemble of velocity fluctuation fields. By introducing an inverse spectral SRT operator the modes can be mapped to velocity fields, facilitating their use as a supplement to conventional TKE-optimized (e-POD) modes. We show that d-POD modes computed from cross sectional slabs of a 3D channel flow DNS ($\Rey_{\tau} = \renum$) reconstruct the dissipation profile more efficiently than do e-POD modes, and vice versa for the TKE profile. 

Although this is to our knowledge the first instance of an explicitly dissipation-optimized POD, several works, in addition to \citet{Lee2020}, have employed optimization with respect to the closely related enstrophy. While distinct quantities, as discussed e.g.\ by \citet{bermejo2009geometry} and \citet{yeung2012dissipation} dissipation and enstrophy remain intimately coupled and display similar spectral properties. They are proportional to the squared norm of the SRT and the vorticity, respectively, both of which are constructed from first-order gradient terms $\nabla^j u^i$. Similar properties with respect to POD convergence may therefore be expected, providing the basis for comparison between results found in the present work and comparable results on enstrophy-based POD found in literature. Enstrophy-based POD was utilized by \citet{huang1994limitations} and by \citet{kostas2005comparison} for identifying coherent flow structures from 2D particle image velocimetry (PIV) measurements of a backward-facing step flow, and similarly by \citet{munir2022proper}, based on 2D measurements of two-phase slug flow made by combining PIV and laser-induced fluorescence. \citet{sengupta2015enstrophy} demonstrated reduced-order modeling of a flow past a cylinder applying enstrophy-based POD to 2D DNS data, using the resulting modes in a vorticity-formulation of the Navier-Stokes equations. Like conventional e-POD modes, enstrophy-based POD modes allow for identification and characterization of important flow structures; the method is of particular use in inhomogeneous flows where vortical structures are of central importance to the dynamics \citep[see][]{sengupta2012instabilities}.

The remainder of this paper is laid out as follows. In section \ref{sec:method} we describe the the general POD formalism and its adaptations in the form of TKE and dissipation optimized PODs employed in this work, we introduce the inverse spectral SRT operator, and we briefly discuss the reconstruction of TKE and dissipation based on the above. 
Implementation details, including the DNS study used to produce the data on which the analysis is built, are discussed in section \ref{sec:channel_flow}. We present a comparison of the lowest POD modes in section \ref{sec:results_modes}. TKE and dissipation are each reconstructed using either basis set, and we analyze the reconstruction of profiles and total quantities in section \ref{sec:results_reconstruction}. Discussion and conclusions are given in sections \ref{sec:discussion} and \ref{sec:conclusion}, respectively.

\section{Proper Orthogonal Decomposition\label{sec:method}}
The POD provides a decomposition of an ensemble of vectors that is optimally efficient as measured by the norm on the underlying vector space \citep[see e.g.][]{holmes2012turbulence,weiss2019tutorial}. A formal and rather generic description of POD is given in section \ref{sec:pod_formalism}. The specifications of the formalism relevant to the energy and dissipation based PODs will be discussed in section \ref{sec:epod_and_dpod}. In section \ref{sec:inverse_srt_op} we introduce the inverse spectral SRT operator used to map d-POD modes to velocity fields. The modal reconstruction of TKE and dissipation is discussed in section \ref{sec:modal_reconstruction}.

\subsection{Generic POD Formalism\label{sec:pod_formalism}}
Let $\mathcal{H}$ be a real Hilbert space of dimension $N$ with inner product $\left(\cdot,\cdot\right)_{\mathcal{H}}: \mathcal{H}\times \mathcal{H}\rightarrow \mathbb{R}$, and let $\alpha \in \mathcal{H}$ be an arbitrary element in $\mathcal{H}$. The induced norm $\lVert\cdot\rVert_{\mathcal{H}}: \mathcal{H}\rightarrow \mathbb{R}_{0+}$ is given by $\lVert \alpha\rVert_{\mathcal{H}} = \left|\left(\alpha, \alpha\right)_{\mathcal{H}}\right|^{\frac{1}{2}}$. Let $\mathcal{F} = \{f_m\}_{m=1}^M \subset\mathcal{H}$ be a set of $M$ samples from $\mathcal{H}$.

Using the averaging operation $\left\langle\cdot\right\rangle$, we define the POD operator $R: \mathcal{H}\rightarrow \mathcal{H}$ for $\mathcal{F}$ by its action on $\alpha$,
\begin{align}
R\alpha  &= \mean{\left(\alpha, f_m\right)_{\mathcal{H}} f_m }{m=1}{M}\,.
\label{eq:pod_operator}
\end{align}

The operator $R$ is Hermitian and has orthogonal eigenvectors $\{\phi_n\}_{n=1}^N$ and real and non-negative eigenvalues $\{\lambda_n\}_{n=1}^N$,
\begin{align}
    R\phi_n &= \lambda_n \phi_n\,, \quad \lambda_n \geq 0\,,\quad \left(\phi_{n},\phi_{n'}\right)_{\mathcal{H}} = \delta_{nn'}\,,
    \label{eq:generic_evp}
\end{align}
where $\delta_{nn'}$ is the Kronecker delta. The eigenvectors are the POD modes, the set of which forms the POD basis, a complete orthogonal basis for $\mathcal{F}$. By convention the POD basis is ordered by decreasing eigenvalues such that $\lambda_1 \geq \lambda_2 \geq \ldots \geq \lambda_N \geq 0$, and the POD modes are normalized, $\left\lVert \phi_n\right\rVert_{\mathcal{H}} = 1$ for $n = 1, 2, \ldots N$. The number of non-zero eigenvalues is $\mathrm{rank}(\mathcal{F}) \leq  \min(M, N)$; in most applications $M \ll N$, and $M$ eigenpairs will be sufficient to completely span $\mathcal{F}$.  For the sake of generality, we will maintain $N$ as our notation for the number of eigenpairs, although in implementations we shall set $N = M - 1$ since the sample mean is subtracted from each sample, reducing the sample set rank by one.

Each sample $f_m\in \mathcal{F}$ can be expanded in the POD basis using coefficients $\{a_{mn}\}_{n=1}^N$,
\begin{align}
    f_m &= \sum_{n=1}^N a_{mn} \phi_n\,, \quad a_{mn} = \left(\phi_n, f_m\right)_{\mathcal{H}}\,.
    \label{eq:pod_expansion_generic}
\end{align}
These coefficients are uncorrelated, satisfying
\begin{align}
    \mean{a_{mn} a_{mn'}}{m=1}{M} &= \lambda_n \delta_{nn'}\,.
    \label{eq:coefficent_correlation}
\end{align}

The optimality of the POD basis with respect to the inner product on $\mathcal{H}$ can be stated formally as
\begin{align}
    \phi_n &= \argmax_{\phi \in \mathcal{H}} \frac{\mean{\left|\left(\phi, f_m\right)_{\mathcal{H}}\right|^2}{m=1}{M}}{\left\lVert\phi\right\rVert_{\mathcal{H}}^2} \quad \text{for}\quad n = 1, 2,\ldots, N\,.
    \label{eq:genereic_optimization}
\end{align}
The eigenvalue problem \eqref{eq:generic_evp} is derived from this optimization problem \citep{Lumley1967}. It results in an optimal basis in the following sense. A sample $f_m\in\mathcal{F}$ may be expanded using the POD basis $\{\varphi_n\}_{n=1}^N$, or alternatively using an arbitrary orthogonal basis $\{\varphi'_n\}_{n=1}^N$ spanning the same subspace of $\mathcal{H}$ as the POD basis. Given $\hat{N} \leq N$ the sample $f_m$ may then be approximated by its truncated expansion in either basis, $\{\phi_n\}_{n=1}^{\hat{N}}$ or $\{\phi'_n\}_{n=1}^{\hat{N}}$:
\begin{align}
    f_m &\approx \hat{f}_m = \sum_{n=1}^{\hat{N}} a_{mn} \phi_n\,;\quad  f_m \approx \hat{f}'_m = \sum_{n=1}^{\hat{N}} a_{mn}' \phi'_n\,.
\end{align}
The POD basis minimizes the mean squared error of the approximation as measured by the norm on $\mathcal{H}$, compared to an arbitrary orthogonal basis,
\begin{align}
    \mean{\lVert \hat{f}_m - f_m\rVert_{\mathcal{H}}^2}{m=1}{M} \leq \mean{\lVert \hat{f}'_m - f_m\rVert_{\mathcal{H}}^2}{m=1}{M}\,,
\end{align}
for any $\hat{N} \leq N$. The sense in which the POD basis is optimal is thus fully determined by the choice of norm $\left\lVert\cdot\right\rVert_{\mathcal{H}}$, and hence by the choice of Hilbert space and inner product.

\subsection{Energy and dissipation based PODs\label{sec:epod_and_dpod}}
We construct two different POD bases using the general formalism laid out in the previous section. One is the conventional TKE-based POD (e-POD), which uses a sample set $\mathcal{U} = \{u_m\}_{m=1}^M \subset \mathcal{H}^{\text{e}}$ of velocity fluctuation snapshots on the spatial domain $\Omega^{\text{e}}$, where the Hilbert space $\mathcal{H}^{\text{e}}$ is
\begin{align}
    \mathcal{H}^{\text{e}} := \left\{\alpha: \Omega^{\text{e}}\rightarrow \mathbb{R}^3 \,\left\lvert\, \sum_{i=1}^3 \int_{\Omega^{\text{e}}} \alpha^i \alpha^i \, dx < \infty \right\}\right.\,.
\end{align}
The inner product on $\mathcal{H}^{\text{e}}$, $(\cdot,\cdot)_{\mathcal{H}^{\text{e}}}:\mathcal{H}^{\text{e}}\times \mathcal{H}^{\text{e}}\rightarrow \mathbb{R}$ is defined as
\begin{align}
    \left(\alpha,\beta \right)_{\mathcal{H}^{\text{e}}} &= \sum_{i=1}^3 \int_{\Omega^{\text{e}}} \alpha^i \beta^i \, dx\,.
\end{align}
The POD basis $\{\varphi_n\}_{n=1}^N$, resulting from solving \eqref{eq:generic_evp}, is optimal with respect  to the mean turbulent kinetic energy (TKE) of the flow, and the associated eigenvalues are $\{\lambda^{\text{e}}_n\}_{n=1}^N$. The mean TKE density, $\left\langle T\right\rangle$, and total mean TKE, $\mathcal{T}$, are
\begin{subequations}
\begin{align}
    \left\langle T\right\rangle &= \frac{1}{2}\mean{\left|u_m\right|^2}{m=1}{M} = \frac{1}{2} \sum_{n=1}^N \lambda_n^{\text{e}} \left|\varphi_n\right|^2\,,\\
    \mathcal{T} &= \int_{\Omega} \left\langle T\right\rangle \, dx = \frac{1}{2}\sum_{n=1}^N \lambda_n^{\text{e}}\,,
\end{align}
\end{subequations}
where $|\alpha|^2 = \sum_{i=1}^3 \alpha^i \alpha^i$ for $\alpha \in \mathcal{H}^{\text{e}}$. This is the commonly used space-only POD, which allows a TKE-optimal reconstruction of the velocity field,
\begin{align}
    u_m &= \sum_{n=1}^N a_{mn} \varphi_n\,,\quad a_{mn}=\left(\varphi_n, u_m\right)_{\mathcal{H}^{\text{e}}}\,.
    \label{eq:vel_expansion}
\end{align}

We propose a second POD variant, namely the dissipation-based POD (d-POD), using instead the ensemble of SRTs computed from the original snapshot ensemble $\mathcal{U}$, i.e., $\mathcal{S} = \{s^{ij}_m\}_{m=1}^M\subset \mathcal{H}^{\text{d}}$ and defined on the spatial domain $\Omega^{\text{d}}$. The Hilbert space $\mathcal{H}^{\text{d}}$ is
\begin{align}
    \mathcal{H}^{\text{d}} &:= \left\{ \alpha : \Omega^{\text{d}}\rightarrow \mathbb{R}^{3\times 3} \left\vert \sum_{i,j=1}^3 \int_{\Omega^{\text{d}}} \alpha^{ij}\alpha^{ij} \, dx < \infty \right\}\right.\,,
\end{align}
with the inner product $(\cdot,\cdot)_{\mathcal{H}^{\text{d}}}:\mathcal{H}^{\text{d}}\times \mathcal{H}^{\text{d}}\rightarrow \mathbb{R}$ defined as
\begin{align}
    \left(\alpha,\beta\right)_{\mathcal{H}^{\text{d}}} &= \sum_{i,j=1}^3 \int_{\Omega^{\text{d}}} \alpha^{ij} \beta^{ij} \, dx\,.
\end{align}

Assuming differentiability of each component of each velocity fluctuation snapshot, $u_m$, the components of the corresponding strain rate tensor, $s_m$, are obtained as $s^{ij}_m \equiv \left(D u_m\right)^{ij}$, with the SRT operator $D: \mathcal{H}^{\text{e}}\rightarrow \mathcal{H}^{\text{d}}$ given by
\begin{align}
    \left(D \alpha\right)^{ij} = \frac{1}{2} \left(\nabla^j \alpha^i + \nabla^i \alpha^j\right)\,.
    \label{eq:srt_operator}
\end{align}

The resulting d-POD modes, $\{\psi_n\}_{n=1}^N$, with associated eigenvalues, $\{\lambda_n^{\text{d}}\}_{n=1}^N$, provide a dissipation-optimal reconstruction of the SRT field,
\begin{align}
    s_m &= \sum_{n=1}^N b_{mn} \psi_n\,, \quad b_{mn} = \left(\psi_n, s_m\right)_{\mathcal{H}^{\text{d}}}\,.
    \label{eq:srt_expansion}
\end{align}

The mean norm on $\mathcal{S}$ is proportional to the mean viscous dissipation of the flow. The mean dissipation density $\left\langle \varepsilon\right\rangle$ and the mean total dissipation $\mathcal{E}$ are given by
\begin{subequations}
\begin{align}
    \left\langle \varepsilon\right\rangle &= 2\nu \mean{\left|s_m\right|^2}{m=1}{M} = 2\nu \sum_{n=1}^N \lambda_n^{\text{d}} \left|\psi_n\right|^2\,, \\
    \mathcal{E} &= \int_{\Omega} \left\langle \varepsilon\right\rangle \, dx = 2\nu \sum_{n=1}^N \lambda_n^{\text{d}}\,,
\end{align}
\end{subequations}
where $|\alpha|^2 = \sum_{i,j=1}^3 \alpha^{ij} \alpha^{ij}$ for $\alpha\in\mathcal{H}^{\text{d}}$.

\subsection{The spectral inverse SRT operator\label{sec:inverse_srt_op}}
An inverse mapping exists, $\mathcal{H}^{\text{d}} \rightarrow \mathcal{H}^{\text{e}}$, mapping SRTs to velocity fluctuation fields. To establish an explicit \textit{spectral} form of this mapping, $D^{-1}$, we consider the subset of d-POD modes for which $\lambda_n^{\text{d}}> 0$, and insert the POD operator definition from \eqref{eq:pod_operator} using $\mathcal{S}$ in the eigenvalue problem \eqref{eq:generic_evp},
\begin{align}
    R\psi_n &= \mean{\left(\psi_n, s_m\right)_{\mathcal{H}^{\text{d}}} s_m}{m=1}{M} = \lambda_n^{\text{d}} \psi_n\,.
\end{align}
Then apply $D^{-1}$ to $\psi_n$ using $D^{-1} s_m = u_m$ to yield
\begin{align}
    D^{-1}\psi_n = \frac{\mean{\left(\psi_n, s_m\right)_{\mathcal{H}^{\text{d}}} u_m}{m=1}{M}}{\lambda_n^{\text{d}}} = \frac{\mean{b_{mn} u_m}{m=1}{M}}{\lambda_n^{\text{d}}}\,.
    \label{eq:inv_srt_mode}
\end{align}

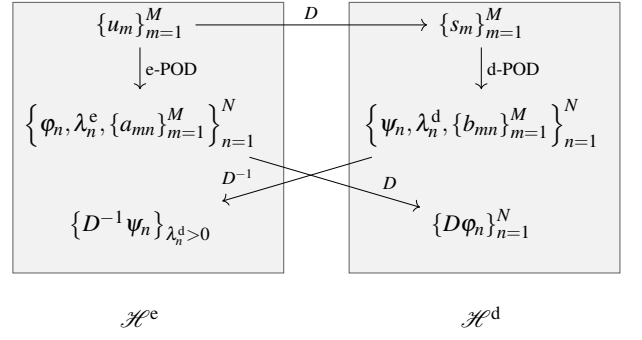
\begin{figure}
    \centering
    \[
        \tikz[overlay]{
            \filldraw[fill=black!5, draw=black!60] (0,-1.4) rectangle ++(3.6,3.6);
            \filldraw[fill=black!5, draw=black!60] (4.47,-1.4) rectangle ++(3.6,3.6);
        }
        \begin{tikzcd}[column sep=large,ampersand replacement=\&]
            \left\{u_m\right\}_{m=1}^M \arrow[r, "D"] \arrow[d, "\text{e-POD}"] \& \left\{s_m\right\}_{m=1}^M\arrow[d, "\text{d-POD}"]\\
            {\left\{\varphi_n, \lambda^{\text{e}}_n, \left\{a_{mn}\right\}_{m=1}^M\right\}_{n=1}^N} \arrow[rd, "D" near end] \& {\left\{\psi_n, \lambda^{\text{d}}_n, \left\{b_{mn}\right\}_{m=1}^M\right\}_{n=1}^N} \arrow[ld, "D^{-1}"' near end] \\
            {\left\{D^{-1}\psi_n\right\}_{\lambda^{\text{d}}_n>0}} \& {\left\{D \varphi_n\right\}_{n=1}^N}\\
            \mathcal{H}^{\text{e}} \& \mathcal{H}^{\text{d}}
        \end{tikzcd}
    \]
    \caption{Schematic depiction of the relation between the objects appearing in the POD. Objects on the left side are associated with the space of velocity fields ($\mathcal{H}^{\text{e}}$), while those on the right are associated with the space of strain rate tensor fields ($\mathcal{H}^{\text{d}}$). The upper row represents samples, the middle row the POD results (modes, eigenvalues, and coefficients), and the lower row the velocity fluctuation fields computed from d-POD modes and the strain rate tensor fields computed from e-POD modes.}
    \label{fig:object_structure}
\end{figure}

The set $\{D^{-1}\psi_n \}_{\lambda^{\text{d}}_n > 0}$ spans the velocity sample space $\mathcal{U}$,
\begin{align}
    u_m &= \sum_{n\mid\lambda^{\text{d}}_n>0} b_{mn} D^{-1}\psi_n\,,
\end{align}
allowing the reconstruction of any velocity-dependent quantity in terms of dissipation-optimized modes. While $\{D^{-1}\psi_n\}_{\lambda^{\text{d}}_n > 0}$ does form a complete basis for $\mathcal{U}$, we note that the basis is in general not orthogonal with respect to the inner product $(\cdot,\cdot)_{\mathcal{H}^e}$. It is, however, orthogonal (and normalized) with respect to the inner product $(\cdot,\cdot)_{\mathcal{H}'}$ defined as
\begin{align}
    (\cdot,\cdot)_{\mathcal{H}'} &= (D\,\cdot,D\,\cdot)_{\mathcal{H}^{\text{d}}}\,,
\end{align}
since $(D^{-1}\psi_n, D^{-1}\psi_{n'})_{\mathcal{H}'} = (\psi_n, \psi_{n'})_{\mathcal{H}^{\text{d}}} = \delta_{nn'}$.
The operator $D$ in \eqref{eq:srt_operator} and its inverse $D^{-1}$ in \eqref{eq:inv_srt_mode} thus link the spaces of velocity fluctuation fields and of SRT fields to each other, as shown schematically in Figure \ref{fig:object_structure}; this link provides useful flexibility when working with the two bases.

\subsection{Modal reconstructions of TKE and dissipation\label{sec:modal_reconstruction}}
By applying the spectral inverse SRT operator the mean TKE and dissipation densities can be reconstructed using \textit{either} basis. We wish to compare the efficiency of these reconstructions in terms of their rate of convergence, leading us to consider the expansions truncated to include $\hat{N} \leq N$ modes. The mean TKE density field reconstructed using $\hat{N}$ e-POD or d-POD modes is
\begin{subequations}
\label{eq:both_mean_tke_density}
\begin{align}
    \left\langle T^{e,\hat{N}}\right\rangle &= \frac{1}{2}\sum_{n=1}^{\hat{N}} \lambda^{\text{e}}_n \left|\varphi_n\right|^2\,,
    \label{eq:epod_mean_tke_density}
    \\
    \left\langle T^{d,\hat{N}}\right\rangle &= \frac{1}{2}\sum_{n=1}^{\hat{N}} \lambda^{\text{d}}_n \left|D^{-1}\psi_n\right|^2\,,
    \label{eq:dpod_mean_tke_density}
\end{align}
\end{subequations}
and the mean dissipation density is likewise reconstructed by
\begin{subequations}
\label{eq:both_mean_dissipation_density}
\begin{align}
    \left\langle \varepsilon^{e,\hat{N}}\right\rangle &= 2\nu \sum_{n=1}^{\hat{N}} \lambda^{\text{e}}_n \left|D\varphi_n\right|^2\,,
    \label{eq:epod_mean_dissipation_density}
    \\
    \left\langle \varepsilon^{d,\hat{N}}\right\rangle &= 2\nu \sum_{n=1}^{\hat{N}} \lambda^{\text{d}}_n \left|\psi_n\right|^2\,.
    \label{eq:dpod_mean_dissipation_density}
\end{align}
\end{subequations}

The e-POD and d-POD reconstructions agree for each quantity when $\hat{N} = N$, but generally exhibit different convergence rates.

\section{Channel flow\label{sec:channel_flow}}
To demonstrate the method presented above we apply it to an ensemble of turbulence fluctuation velocity data $\{u_m\}_{m=1}^M$ obtained from a DNS of a double-periodic channel flow. This is chosen as an example of a relatively simple and well-understood flow, while still exhibiting shear layers and inhomogeneous turbulence which highlight relevant features of the method. This section presents computational details of the DNS and the subsequent data processing.

The DNS is performed using the Ellipsys3D code \citep[see][]{michelsen1992basis3d,michelsen1994block,sorensen1995}. EllipSys3D is a finite volume code that solves the incompressible Navier--Stokes equations in general curvilinear coordinates on a collocated grid arrangement. Time stepping is performed with a second-order accurate implicit method, which uses subiterations to converge residuals within each time step. The pressure correction equations are solved using a SIMPLEC-like algorithm with Rhie/Chow interpolation to prevent odd/even pressure decoupling.
This combines the ideas of \citet{Kobayashi1991}, which use the non-relaxed momentum ($\alpha_u = 1$) equations for deriving the interface fluxes and the pressure correction equation, with the approach of \citet{Shen2003}, in which the Rhie/Chow fluxes from the previous time steps are used to account for the corresponding velocities.

The flow domain has streamwise, transverse, and spanwise dimensions $L_x\times L_y\times L_z = 4\pi\times2\times2\pi$, or in wall units, $L_x^+\times L_y^+ \times L_z^+ = \num{4906}\times \num{781}\times \num{2453}$. The flow is resolved on a regular grid of $\thenx\times \theny \times \thenz = \num{\thengrid}$ points which is homogeneous in the streamwise and spanwise directions. The wall normal discretization follows a hyperbolic sine, which concentrates points near the walls ($\Delta y_{\text{w}} \approx \num{6.8e-5}$) and gradually coarsens to $\Delta y_{\text{c}} \approx \num{0.011}$ at the center of the channel, corresponding to $\Delta y_{\text{w}}^+ \approx 0.027$ and $\Delta y_{\text{c}}^+ \approx 4.3$. The spatial resolution is comparable to the recommended values by \cite{YangDNS2021}. The time step is $\Delta t = \num{5e-3}$ corresponding to a CFL condition of approximately $0.20$ in the center of the channel.

The channel is periodic in the spanwise and streamwise directions, and bounded by no-slip boundary conditions in the transverse direction. Constant mass flux is enforced with a PID-controller, which ensures faster flow convergence compared to applying a constant pressure gradient. The simulation setup is summarized in Table \ref{tab:domain_grid_bc}.

\begin{table}[ht]
    \centering
    \caption{Boundary conditions and domain and grid parameters. $\Delta y_\text{w}$ and $\Delta y_{\text{c}}$ designate the transverse resolution at the wall and at the center of the channel, respectively.\label{tab:domain_grid_bc}}
    \begin{ruledtabular}
    \begin{tabular}{c|c|r@{ $=$}l|c}
         &  Streamwise, $x$ & \multicolumn{2}{c|}{Transverse, $y$} & Spanwise, $z$ \\\hline
         Bound.\ cond. & Periodic & \multicolumn{2}{c|}{No slip} & Periodic \\
         && \multicolumn{2}{c|}{} &\\
         Domain size && \multicolumn{2}{c|}{} &\\
         (flow units) & $4\pi$ & \multicolumn{2}{c|}{$2$} & $2\pi$\\
         (wall units) & $\num{4906}$ & \multicolumn{2}{c|}{$\num{781}$} & $\num{2453}$\\
          && \multicolumn{2}{c|}{} &\\
         Grid points & $\thenx$ & \multicolumn{2}{c|}{$\theny$} & $\thenz$ \\
         && \multicolumn{2}{c|}{} &\\
         Resolution && \multicolumn{2}{c|}{} &\\
         (flow coord.) & $\num{0.028}$ & $\Delta y_{\text{w}}$&$ \num{6.8e-5}$ & $\num{0.016}$ \\
          && $\Delta y_{\text{c}}$&$ \num{0.011}$ & \\
         (wall coord.) & $11.0$ & $\Delta y_{\text{w}}^+$&$ \num{0.027}$ & $\num{6.4}$ \\
          && $\Delta y_{\text{c}}^+$&$ \num{4.3}$ & \\
          %&& $\Delta y_{\text{c}}^+$&$ \num{4.432}$ & \\
    \end{tabular}
    \end{ruledtabular}
\end{table}

The bulk and friction Reynolds numbers are, respectively,
\begin{align}
    \Rey &= \frac{U_0 h}{\nu} \approx 8800\,,\quad  \Rey_{\tau} = \frac{u_{\tau} h}{\nu} \approx \renum\,,% \quad u_{\tau} = \sqrt{\nu \left(\frac{dU}{dy}\right)_{\text{wall}}}
\end{align}
where $U_0$ is the spanwise average of the mean streamwise center line velocity, $u_{\tau}$ is the friction velocity, %$U$ the spanwise average mean velocity, $y$ the wall normal coordinate, 
$h=1$ the channel half-width, and $\nu$ is the kinematic viscosity. A second order central difference gradient stencil is used for obtaining the derivative of the average streamwise velocity profile $U$ at the wall.

Velocity fields are extracted on a cross-sectional slab with a streamwise width of three grid points, giving three neighboring planes ($p_1$, $p_2$, and $p_3$) of $\theny\times\thenz = \num{\thenplane}$ points each. This allows for computation of all gradient components in the central plane $p_2$ using a second order central difference stencil. The entire slab $\Omega^{\text{e}} = \cup_{i=1}^3 p_i$ is used as domain for computing e-POD modes, to facilitate subsequent computation of modal SRTs, whereas SRT elements derived from the gradients computed on $\Omega^{\text{d}} = p_2$ are used for d-POD modes. The resulting dimension of e-POD modes, with three independent components, is therefore $3|\Omega^{\text{e}}| = 3\times 3\times\theny\times\thenz = \num{\thenepoddom}$, and for d-POD modes with six independent components, $6|\Omega^{\text{d}}| = 6\times\theny\times\thenz = \num{\thendpoddom}$.

\begin{figure}
    \centerline{\input{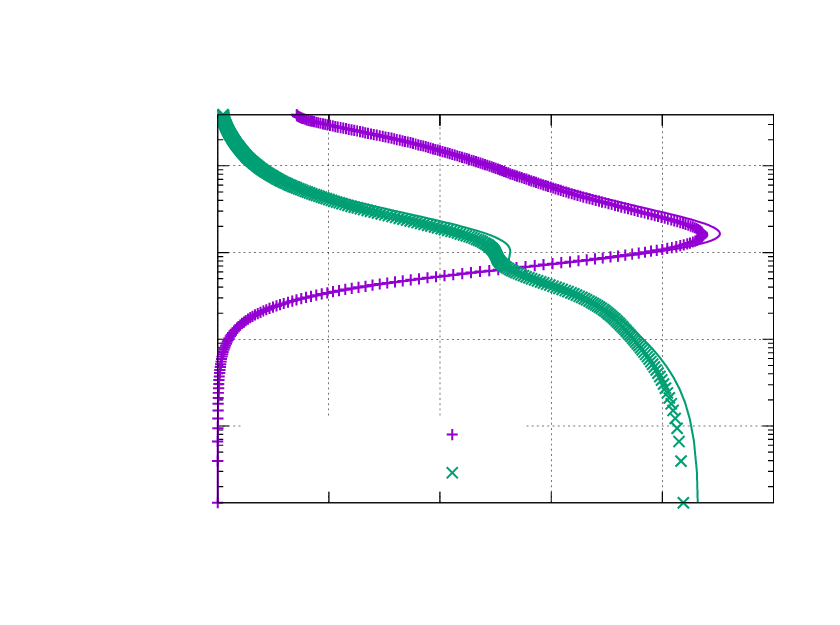}}
    \caption{Transverse profiles of TKE (lower axis) and dissipation (upper axis). Symbols indicate values obtained in this work, and solid lines values obtained by \citet{iwamoto2002database,iwamoto2002reynolds}.\label{fig:tke_diss_prof}}
\end{figure}

Transverse profiles of mean TKE and dissipation obtained from the snapshot samples are shown in Figure \ref{fig:tke_diss_prof}, along with profiles obtained by \citet{iwamoto2002database,iwamoto2002reynolds} for comparison. The overall profiles are reproduced qualitatively, although we find TKE to be slightly underestimated near the TKE maximum at $y^+ \approx 20$, and likewise dissipation around $y^+ \approx 10$ and $y^+\lesssim 1$. However, these discrepancies are minor and should not affect the conclusions drawn in this work. Profiles of mean velocity, Reynolds shear and normal stresses, and TKE production are all in excellent agreement with the results of \citet{iwamoto2002database,iwamoto2002reynolds}, and are not shown.

To approximate independent sampling the fluctuation velocity snapshot ensemble is formed by extracting realizations at an interval of 100 time steps. The integral time scale $T_E$ calculated at the center of the channel corresponds to 145 time steps, corresponding to a separation between samples of $0.69 T_E$. This implies that the samples are not expected to be fully uncorrelated. The total number of time steps in the simulation from which samples are extracted is $\num{\theallsamples}$, and we obtain a total of $M = \num{\thensamples}$ samples. Since the mean field is subtracted, the dimension of the POD spaces is \addtocounter{nsamples}{-1} $N = M-1 = \thensamples$.

The POD is realized by performing a singular value decomposition (SVD) for the matrix $\widetilde{F}$, which is the $d|\Omega|\times M$ data matrix whose columns are snapshot vectors, appropriately weighted to implement the quadrature rule for the grid. Here, $d=3$ and $\Omega = \Omega^{\text{e}}$ for the e-POD, while $d=6$ and $\Omega=\Omega^{\text{d}}$ for the d-POD. The weighting amounts to a similarity transform which is subsequently inverted for the modes produced.

The SVD is computed with the SLEPc computational toolkit SVD solver \citep[see][]{Balay1997,fahl2001computation,petsc-user-ref,roman2022slepc}. The solver implements the SVD using the cross product method, in which the eigenvalue problem is solved for the matrix $\widetilde{\mat{F}}^T\widetilde{\mat{F}}$ and the resulting eigenvectors are mapped to POD eigenmodes (left singular vectors) and coefficients (right singular vectors), while POD eigenvalues are obtained from the singular values. This procedure is equivalent to the snapshot POD algorithm described by \citet{sirovich1987turbulence1}. As $M \ll d |\Omega|$ this approach reduces the memory cost of the problem when compared to the more direct approach of solving the eigenvalue problem for $\widetilde{F}\widetilde{F}^T$.

Due to the spanwise periodicity and homogeneity of the flow the spanwise part of the POD modes can be shown analytically to be harmonic functions. This can be utilized by employing the spectral POD (SPOD), in which the spanwise Fourier transform of the autocorrelation matrix is decomposed separately for each wave number. This method was introduced by \citet{Lumley1967} for eliminating homogeneous directions from the POD eigenvalue problem, and applied to the azimuthal direction of an axisymmetric jet by \citet{citriniti2000reconstruction}. The application of SPOD has the potential to dramatically reduce the dimensionality of the POD eigenvalue problem compared to non-spectral POD (or physical POD, in the nomenclature of \citet{picard2000pressure}). However, it may also complicate the analysis and interpretation of the resulting modes, as SPOD modes are indexed by both wave number and mode number, rather than by mode number alone. Furthermore, reconstructing physical fields from the modes is complicated by the loss of phase information \citep{aubry1988dynamics}. Among the goals of the present work is to gain insight into the nature of modes obtained using d-POD compared to e-POD, and in order to avoid complications arising from the use of SPOD we apply only the physical POD here. However, we do note that valuable insights could also be gained from the SPOD analysis, pertaining in particular to the spectral properties of modes.

\section{Energy and Dissipation Modes\label{sec:results_modes}}
\begin{figure*}
    \centerline{\input{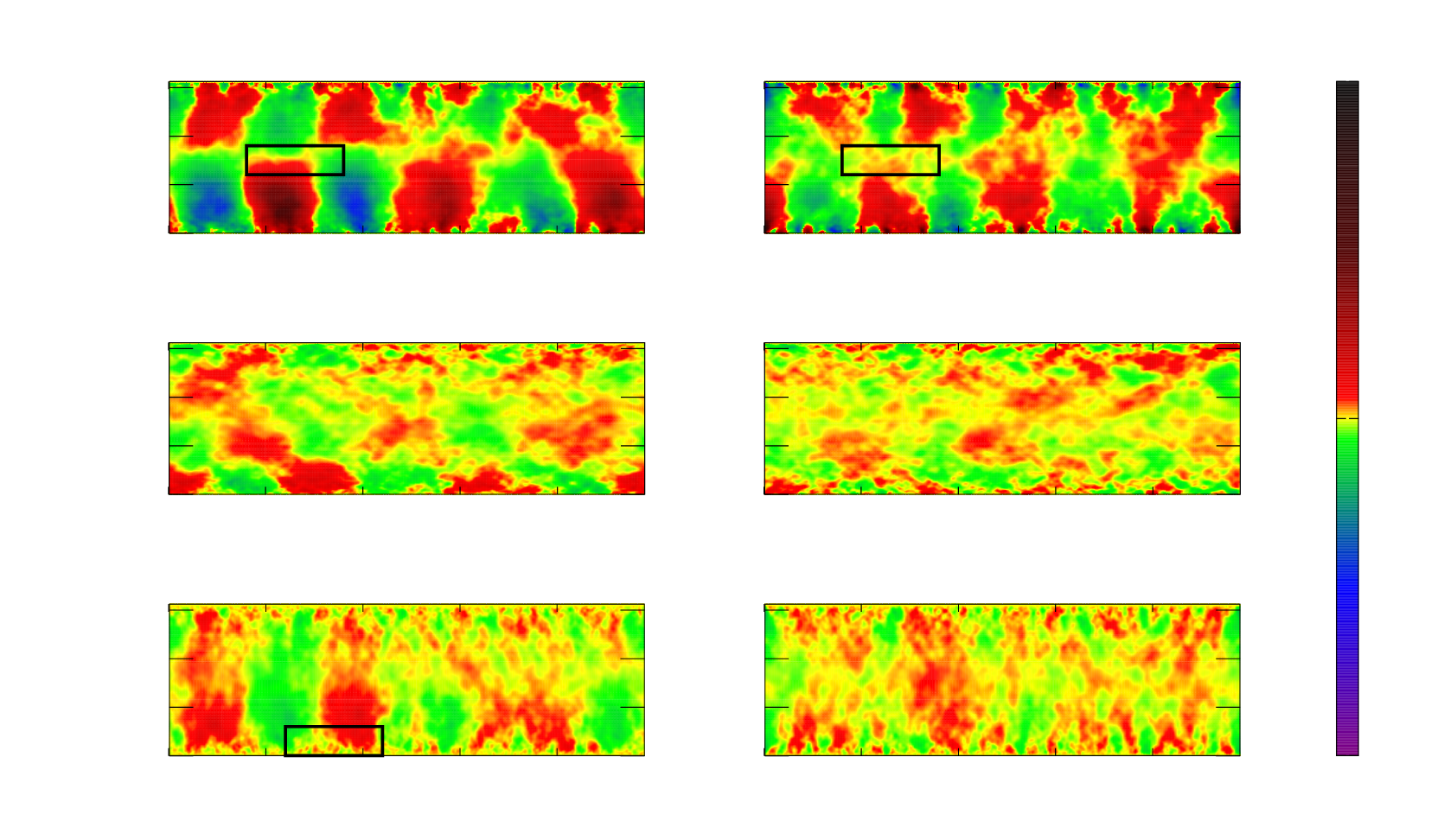}}
    \caption{\emph{Left:} Components $\varphi^i_1$ of the lowest e-POD mode. \emph{Right:} Components $(\widetilde{D^{-1}\psi_1})^i$ of the normalized velocity field corresponding to the lowest d-POD mode. Boxes mark regions shown in enlarged view in Figure \ref{fig:mode_zooms}. Axes show coordinates in wall units (spanwise horizontal, transverse vertical), and the color scale shows values with the full range normalized to $[-1;1]$.
    \label{fig:epod_and_predpod_mode}}
\end{figure*}

In order to gain a sense of the relation of the two bases, components of the lowest e-POD mode, $\varphi_1$, are compared to the components of $\widetilde{D^{-1} \psi}_1$ which is the normalized velocity fluctuation field corresponding to the lowest d-POD mode, obtained using \eqref{eq:inv_srt_mode}. The three components of the lowest e-POD mode in the central plane $p_2$ are shown on the left in Figure \ref{fig:epod_and_predpod_mode}, with the components of the velocity field corresponding to the lowest d-POD mode shown on the right in Figure \ref{fig:epod_and_predpod_mode}.

\begin{figure}
    \centerline{\input{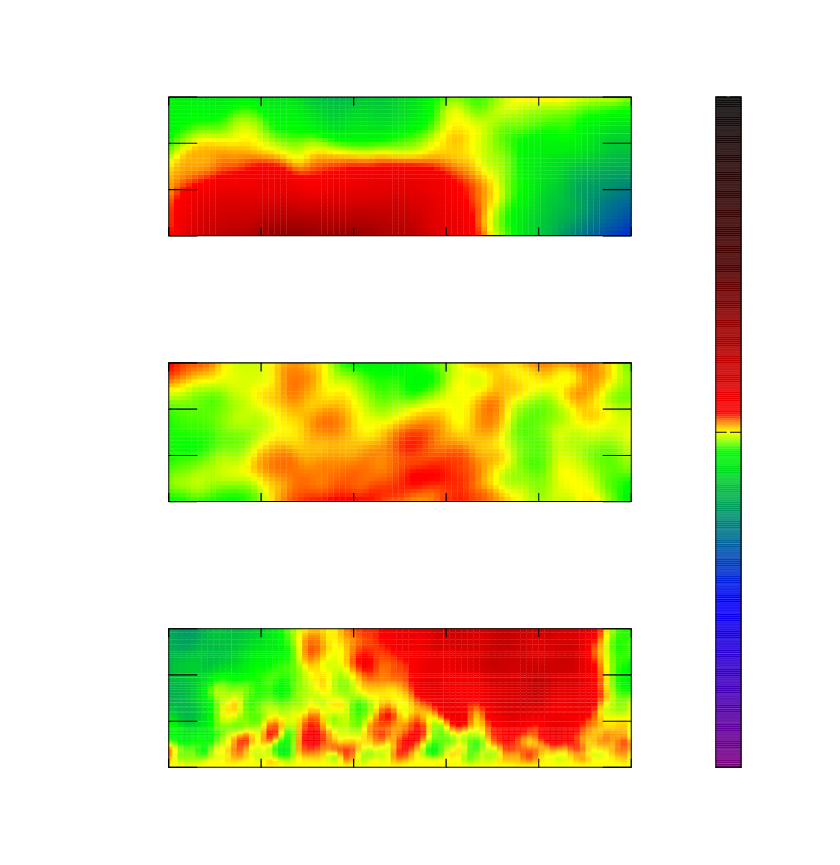}}
    \caption{Enlarged view of regions marked with dashed boxes in Figure \ref{fig:epod_and_predpod_mode}. $(a)$: section in the bulk in $\varphi_1^1$. $(b)$: section at the same position in $(D^{-1}\psi_1)^1$. $(c)$: Section near the lower wall in $\varphi_1^3$.
    For explanation of axes and color scale, see the caption to Figure \ref{fig:epod_and_predpod_mode}.
    \label{fig:mode_zooms}}
\end{figure}

The e-POD mode and the d-POD velocity fields shown in Figure \ref{fig:epod_and_predpod_mode} both exhibit structures across a range of scales, and somewhat similar large scale structures can be recognized among the two. The d-POD velocity field shows a richer small-scale structure than that found in the e-POD mode; this is most clearly seen in the streamwise component, as emphasized in the enlarged views of $\varphi_1^1$ and $(\widetilde{D^{-1}\psi}_1)^1$ in Figure \ref{fig:mode_zooms}a and \ref{fig:mode_zooms}b, respectively. In both the e-POD mode and the d-POD velocity field, large scale structures are more prominent and well-defined in the streamwise components, where the magnitude of turbulent fluctuations are generally larger, while more structures on smaller scales are apparent in the transverse and spanwise components. Both exhibit richer small scale structure near the walls ($|y^+ - y_{\text{w}}^+| \lesssim 50$, where $y_{\text{w}}^+ \in \{0, 784\}$ denotes the wall locations) than in the bulk. Examples of near-wall small scale structures in $\varphi_1^3$ are found in the lower part of Figure \ref{fig:mode_zooms}c. The trends outlined here are also found in the next several e-POD modes and d-POD velocity fields which are not shown.

\begin{figure*}
    \centerline{\input{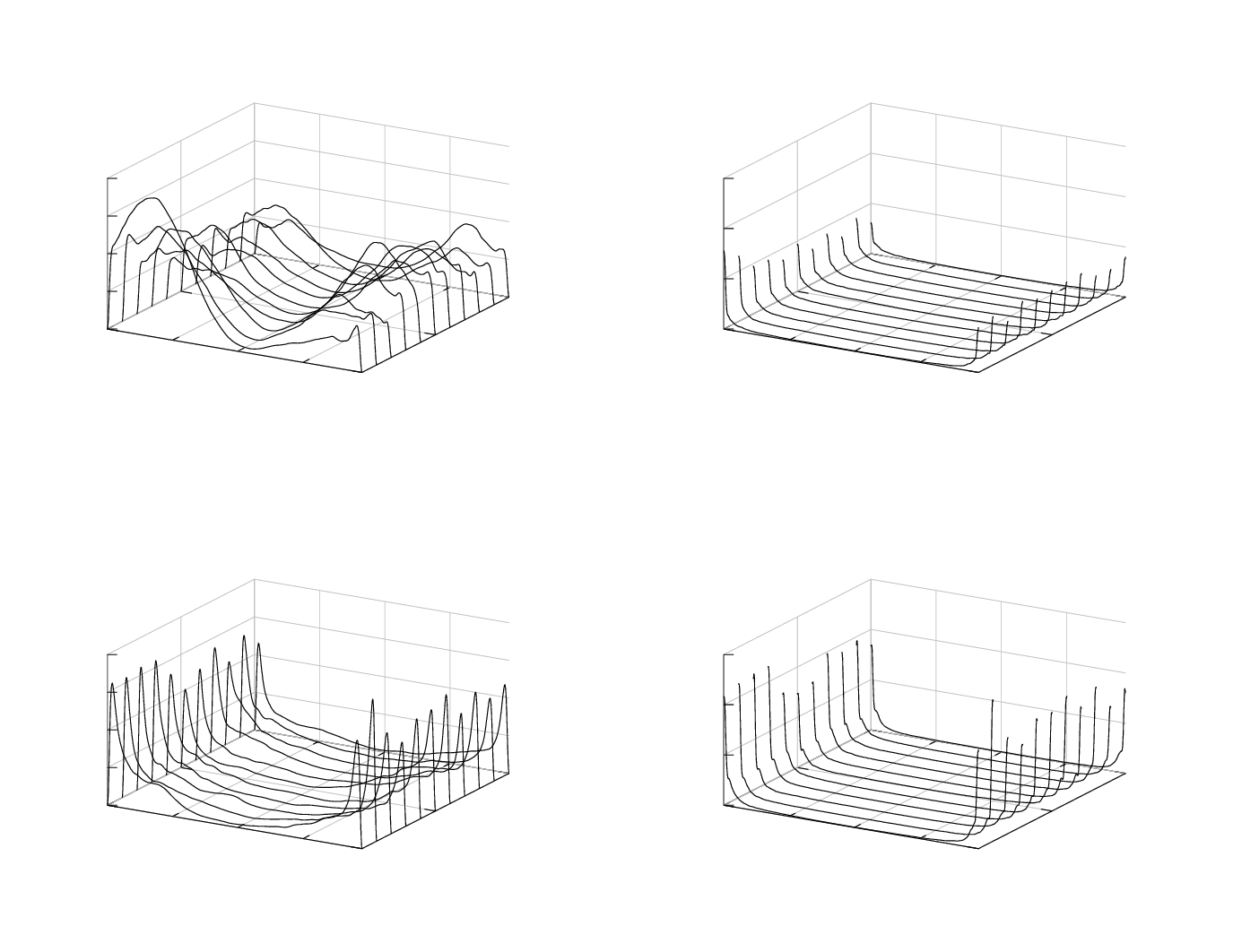}}
    \caption{Contributions of the 10 lowest e-POD modes $(a)$ and d-POD modes $(b)$ to the mean TKE profile, of the same modes ($(c)$ and $(d)$, respectively) to the dissipation profile.\label{fig:lineplot_reconstructions}}
\end{figure*}

Transverse TKE and dissipation profiles are reconstructed from both decompositions, producing a total of four reconstructions. The contributions of each of the lowest several modes of either decomposition to the reconstruction of transverse profiles of TKE and dissipation are shown in Figure \ref{fig:lineplot_reconstructions} (the full reconstructions are considered in the next section).
The transverse profiles are obtained as spanwise averages of the corresponding terms in the sums in \eqref{eq:both_mean_tke_density} for TKE and \eqref{eq:both_mean_dissipation_density} for dissipation. Here, the differences between the two decompositions become clearer; while the TKE contributions of the lowest several e-POD modes are distributed across the domain, with a dip around the TKE-poor center of the channel, the corresponding d-POD mode TKE contributions all peak near the walls and decrease strongly towards the bulk. For either type of modes, dissipation contributions are concentrated close to the wall; however, the contribution from the lowest d-POD modes is much larger than that of the corresponding e-POD modes.

Although the two decompositions exhibit similar large-scale structures as seen in Figure \ref{fig:epod_and_predpod_mode}, important differences of the decompositions are revealed in Figure \ref{fig:lineplot_reconstructions}. The fact that similarities show up in the first place may be explained in part by the underlying dynamics, and in part by the applied method. As to the former, the moderate Reynolds number implies a narrow inertial range, leading to less scale separation between energetic and dissipative structures, thus producing non-negligible overlap between these. As to the latter, since POD modes are purely empirical and reflect the underlying dynamics of the flow only to the extent that it is represented in the sample of snapshots, the sample size is a crucial parameter. The snapshots sampled here produce snapshot matrices of full rank, implying that the sampling does not capture the full extent of the phase space. The effect of this is more pronounced near the walls, where the low streamwise mean velocity leads to weaker advection and longer coherent time scales. This may potentially cause an overlap between TKE-rich and dissipative structures present, contributing to similarities in the modes obtained from the two decompositions.

\section{TKE and dissipation reconstruction\label{sec:results_reconstruction}}
Both TKE and dissipation may be reconstructed using either e-POD modes or d-POD modes, as described in Section \ref{sec:modal_reconstruction}. These reconstructions are shown and analyzed in Section \ref{sec:profile_reconstruction}. The relative efficiencies of the reconstructions are quantified in terms of the reconstruction efficiency gain, which is introduced and analyzed in Section \ref{sec:reconstruction_efficiency_gain}. The convergence of integrated TKE and dissipation are analyzed in Sections \ref{sec:integrated_tke_convergence} and \ref{sec:integrated_dissipation_convergence}, respectively. Finally, we consider the inherent TKE and dissipation content in d-POD and e-POD modes in Section \ref{sec:ensmc}.

\subsection{Profile reconstructions\label{sec:profile_reconstruction}}
\begin{figure*}
    \centerline{\input{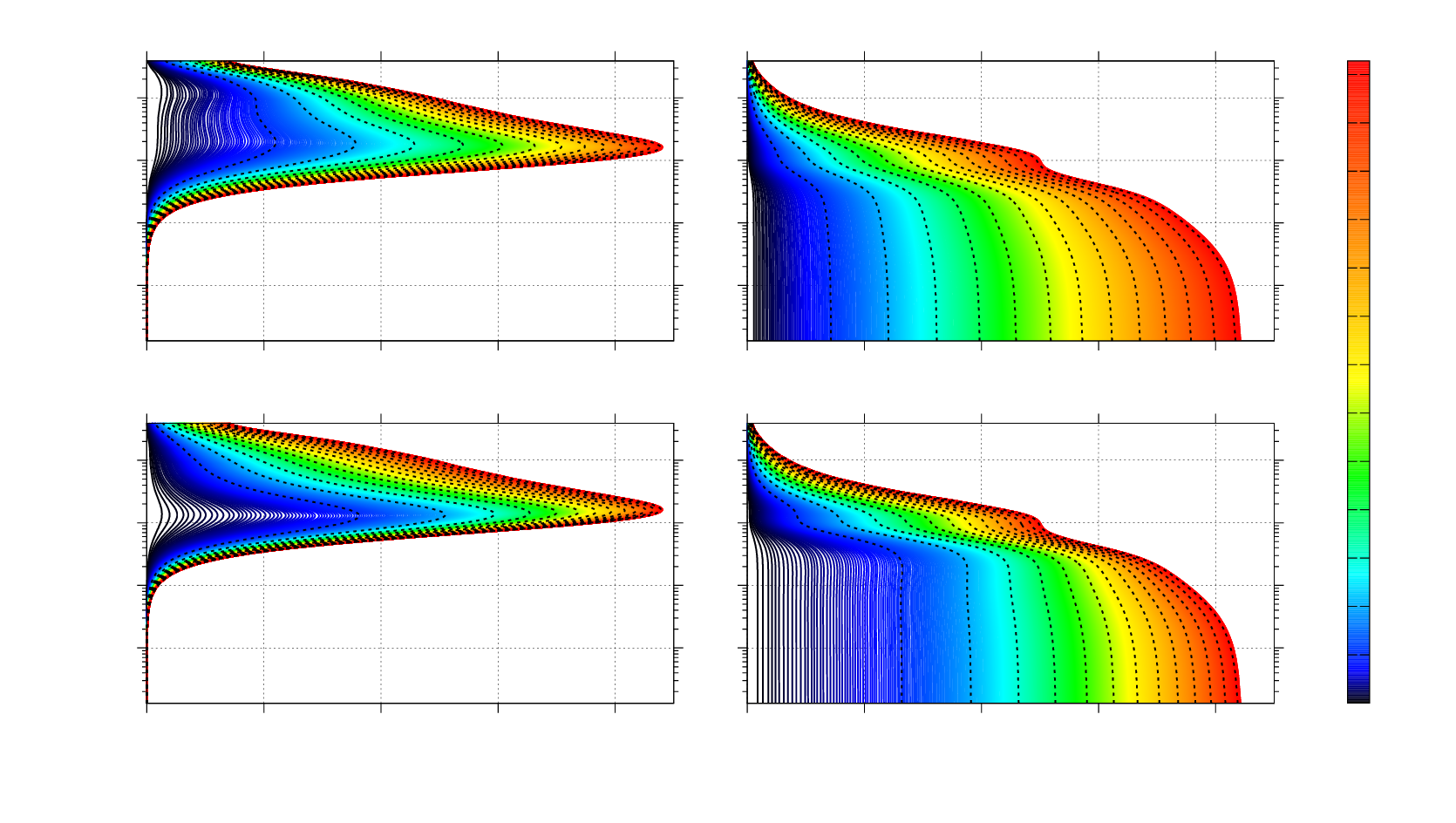}}
    \caption{Reconstruction of TKE profile using e-POD modes ($a$) and d-POD modes ($b$), and of dissipation profile using e-POD modes ($c$) and d-POD modes ($d$). Colors show the profile reconstructed using a corresponding number of e-POD (($a$) and ($c$)) or d-POD modes (($b$) and ($c$)). Dashed lines trace reconstructed profiles for each 50 additional modes.}
    \label{fig:all_incremental_profiles}
\end{figure*}

Partial reconstructions of the transverse TKE and dissipation profiles are performed by varying the number of terms $\hat{N}$ included in the expansions in \eqref{eq:both_mean_tke_density} for TKE and \eqref{eq:both_mean_dissipation_density} for dissipation, and averaging in the spanwise direction and across the channel center line. Figure \ref{fig:all_incremental_profiles} shows reconstructions of TKE and dissipation profiles using e-POD and d-POD modes. For each additional mode included in the reconstruction the resulting profile is shown as a colored line, with line colors indicating the number of modes included. Dashed contour lines mark the profiles reconstructed for each additional 50 modes, corresponding to the tic markings on the color bar. The horizontal distance between successive contours at a given position $y^+$ indicates the additional TKE (Figures \ref{fig:all_incremental_profiles}a and \ref{fig:all_incremental_profiles}b) or dissipation density (Figures \ref{fig:all_incremental_profiles}c and \ref{fig:all_incremental_profiles}d) contributed to the reconstruction by the corresponding 50 modes. Larger horizontal jumps between dashed lines and a more spread out color gradient are thus indicative of a greater per-mode contribution at the given position and stage of the reconstruction. A more efficient reconstruction is indicated by a more rapid build-up in the initial part of the reconstruction, pushing the color gradient further to the right. For the lowest modes, the accumulated contributions from individual modes are discernible as discrete profiles.

Figure \ref{fig:all_incremental_profiles}a shows that fewer than 50 e-POD modes are needed to reproduce the TKE maximum around $y^+ \approx 20$, although the profile becomes similar in shape to the full profile only at larger numbers of modes. The leftmost contours are well spaced, and the rightmost more densely placed; this indicates that the contribution per mode to the TKE profile is much larger for lower modes than for higher modes, demonstrating the efficiency of the reconstruction.

Compared to the e-POD TKE reconstruction, TKE is reconstructed remarkably efficiently near the wall  using d-POD modes, including the region around the TKE maximum, as shown in Figure \ref{fig:all_incremental_profiles}b. This efficiency does not extend into the bulk where most of the TKE is located. As suggested by Figures \ref{fig:lineplot_reconstructions}a and \ref{fig:lineplot_reconstructions}b, the d-POD prioritizes the near-wall region at the expense of the bulk, due to dissipation being concentrated in the former. This enables efficient TKE reconstruction using d-POD modes in the near-wall region, but not in the bulk.

Figure \ref{fig:all_incremental_profiles}c shows the e-POD dissipation profile reconstruction. There is little difference in the spacing of contour lines between lower and higher modes in the dissipation-rich near-wall region, indicating that modal dissipation contributions are comparable throughout a large part of the spectrum, and that e-POD modes are poorly optimized to reconstruct dissipation.

The d-POD dissipation profile reconstruction is shown in Figure \ref{fig:all_incremental_profiles}d. In the near-wall region we find a significant fraction of the profile reconstructed with relatively few modes; contour lines for higher mode numbers lie more closely near the full profile outline. This implies the expected more efficient dissipation reconstruction using d-POD than e-POD. As also suggested by Figure \ref{fig:lineplot_reconstructions}b, the d-POD prioritizes the TKE-poor but dissipation-rich wall region, leading to higher efficiency in reconstructing this dissipation-rich part of the dissipation profile. A more modest difference between e-POD and d-POD dissipation reconstruction is observed in the dissipation-poor region further from the wall.

\subsection{Reconstruction efficiency gain\label{sec:reconstruction_efficiency_gain}}
A more systematic comparison of reconstruction efficiencies can be made by comparing the number of modes of either decomposition needed to reconstruct a given fraction of a reconstructed profile. We thus define the \textit{reconstruction efficiency gain} (REG) for TKE and for dissipation as
\begin{subequations}
\label{eq:num_gain}
\begin{align}
    g_T(f_T, y^+) &= \frac{N_{\text{e}, T}(f_T, y^+)}{N_{\text{d}, T}(f_T, y^+)}
    \label{eq:num_gain_T}\,,\\
    g_{\varepsilon}\left(f_{\varepsilon}, y^+\right) &= \frac{N_{\text{e}, \varepsilon}\left(f_{\varepsilon}, y^+\right)}{N_{\text{d}, \varepsilon}\left(f_{\varepsilon}, y^+\right)}\,,
    \label{eq:num_gain_E}
\end{align}
\end{subequations}
where $N_{\text{e},T}(f_T, y^+)$ and $N_{\text{d},T}(f_T, y^+)$ are the number of e-POD and d-POD modes, respectively, needed to reconstruct the fraction $f_T$ of the local mean TKE at position $y^+$, and $N_{\text{e},\varepsilon}(f_{\varepsilon}, y^+)$ and $N_{\text{d},\varepsilon}(f_{\varepsilon}, y^+)$ denote the corresponding quantities for dissipation instead of TKE.

A larger REG signifies that fewer modes are needed in the d-POD reconstruction compared to the e-POD reconstruction when reconstructing the fraction $f_{T}$ of the TKE profile (for $g_T$), or the fraction $f_{\varepsilon}$ of the dissipation profile (for $g_{\varepsilon}$), at position $y^+$. The REG thus quantifies the relative lead in local convergence of d-POD over e-POD for a given fraction of TKE or dissipation. As the fraction $f_T$ or $f_{\varepsilon}$ approaches unity, the reconstruction comes to include all modes in either basis, and the REG also approaches unity.

\begin{figure}
    \centerline{\input{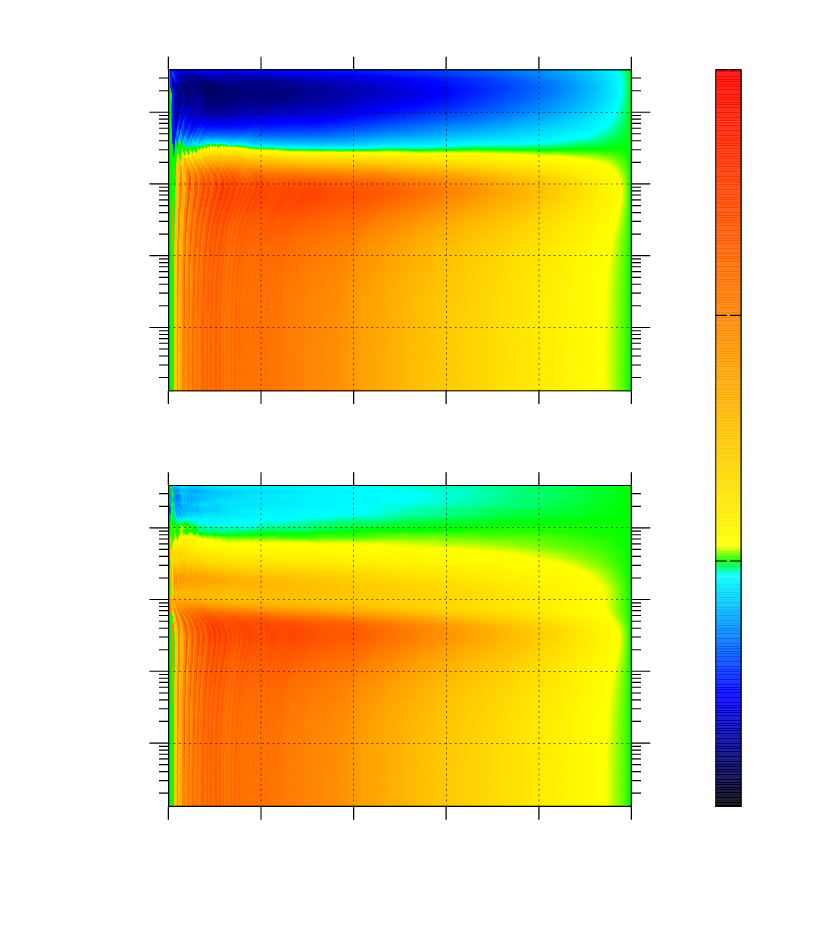}}
    \caption{Reconstruction efficiency gain (REG) as defined in \eqref{eq:num_gain}. $(a)$: TKE REG $g_T$ as a function of TKE fraction $f_T$ and position $y^+$. $(b)$: Dissipation REG $g_{\varepsilon}$ as a function of dissipation fraction $f_{\varepsilon}$ and position $y^+$.\label{fig:num_gain}}
\end{figure}

Computed REGs are shown in Figure \ref{fig:num_gain}. We note that both REGs tend to approach unity approximately monotonically when increasing the fraction $f_{T}$ or $f_{\varepsilon}$; in particular, whether either REG is above or below unity is mainly determined by the wall coordinate. Both REGs exhibit values larger than unity near the wall, implying that the d-POD reconstructions of both TKE and dissipation are more efficient than the corresponding e-POD reconstructions in this region, while the reverse is true in the bulk. The region where $g_{\varepsilon} > 1$ extends further into the bulk ($y^+ \lesssim 70$) than the region where $g_T > 1$ ($y^+ \lesssim 30$), reflecting the optimality of d-POD with respect to the reconstruction of dissipation. We also note that $g_T$ tends to attain more extreme values compared to $g_{\varepsilon}$, particularly for $y^+ \gtrsim 8$, indicating smaller differences in dissipation reconstruction efficiency using the d-POD compared to the e-POD. 

Figure \ref{fig:num_gain}a shows the TKE REG, $g_T$. The very low TKE REG in the bulk region supports the conclusion drawn from Figures \ref{fig:all_incremental_profiles}a and \ref{fig:all_incremental_profiles}b that the TKE-rich bulk structures are most efficiently reconstructed using e-POD, while the same structures are given low priority in the d-POD due to their low dissipation contribution. The TKE maximum ($y^+ \approx 20$) is located near the edge of the region in which TKE REG is above unity, but most of the TKE is localized within the low-REG bulk region with appreciable TKE density, which covers a much larger volume (although appearing compressed in Figure \ref{fig:num_gain} due to logarithmic scaling). The large TKE REG near the wall reflects the efficiency of d-POD TKE reconstruction in this region observed in Figures \ref{fig:lineplot_reconstructions}b and \ref{fig:all_incremental_profiles}b, due to the prevalent dissipative structures also carrying a significant part of what little TKE is contained in this region. As these structures contribute little to the total TKE they are given low priority in the e-POD reconstruction, whereas they contribute much of the total dissipation, giving them a high priority in the d-POD reconstruction.

The dissipation REG, $g_{\varepsilon}$, is plotted in Figure \ref{fig:num_gain}b and exhibits the same overall patterns as those seen in $g_{T}$. It shows above-unity values near the wall, confirming the enhanced d-POD dissipation reconstruction efficiency in the region seen in Figure \ref{fig:all_incremental_profiles}d. The value increases when moving closer to the wall, peaks around $y^+ \approx 8$, and remains high. The high dissipation REG very close to the wall corresponds the region where TKE is suppressed, causing the e-POD to give very low priority to structures localized here. Conversely, the moderately sub-unity values of $g_{\varepsilon}$ in the far bulk ($y^+ \gtrsim 100$) reflect the very limited dissipation contributed by structures in this region, which causes the d-POD to not prioritize these structures. Because of their appreciable TKE contribution they are reconstructed more efficiently by e-POD, and their dissipation contribution is therefore included fairly efficiently in the e-POD reconstruction. This difference is modest, however, causing the dissipation REG to remain near unity even in the bulk.

\subsection{Convergence of integrated TKE \label{sec:integrated_tke_convergence}}
We consider the integrated and normalized partial TKE reconstructions, $\widetilde{\mathcal{T}}^{\text{e},\hat{N}}$ and $\widetilde{\mathcal{T}}^{\text{d},\hat{N}}$, obtained from integrating the TKE reconstructed with only the first $\hat{N}$ e-POD modes using \eqref{eq:epod_mean_tke_density}, or with only the first $\hat{N}$ d-POD modes using \eqref{eq:dpod_mean_tke_density}, respectively, and normalizing by the value for the full TKE, $\mathcal{T}^N$:
\begin{align}
    \mathcal{T}^{\text{e/d},\hat{N}} &= \int_{\Omega} \left\langle T^{\text{e/d},,\hat{N}}\right\rangle \, dx\,,\quad \widetilde{\mathcal{T}}^{\text{e/d},\hat{N}} = \frac{\mathcal{T}^{\text{e/d},\hat{N}}}{\mathcal{T}^{N}}\,
\end{align}

The integrated partial TKEs for e-POD and d-POD are, respectively,
\begin{subequations}
\begin{align}
    \mathcal{T}^{\text{e},\hat{N}} &= \frac{1}{2}\sum_{n=1}^{\hat{N}} \lambda^{\text{e}}_n\,,
    \label{eq:tke_epod_integrated}
    \\
    \mathcal{T}^{\text{d},\hat{N}} &= \frac{1}{2}\sum_{n=1}^{\hat{N}} \lambda^{\text{d}}_n \left\lVert D^{-1} \psi_n\right\rVert_{\mathcal{H}^{\text{e}}}^2\,.
    \label{eq:tke_dpod_integrated}
\end{align}
\end{subequations}

\begin{figure}
    \centerline{\input{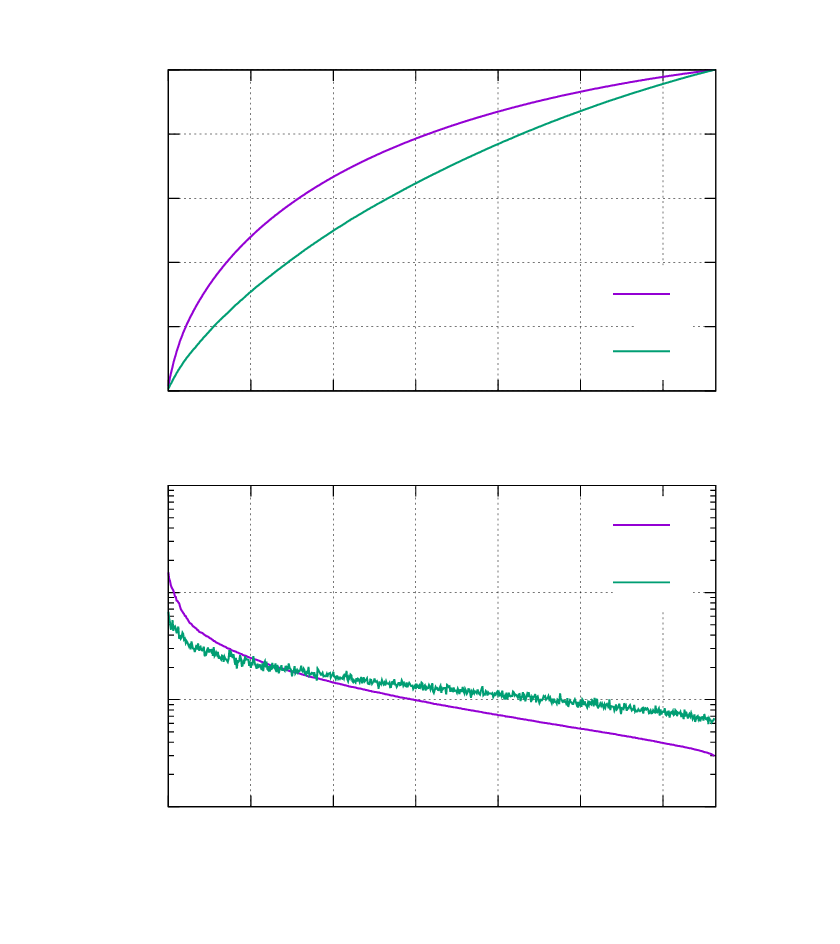}}
    \caption{$(a)$: Convergence of normalized integrated partial TKE reconstructions $\widetilde{\mathcal{T}}^{\text{e/d},\hat{N}}$ vs number $\hat{N}$ of e-POD and d-POD modes. $(b)$: Single mode TKE contribution $\Delta\widetilde{\mathcal{T}}^{\text{e/d},\hat{N}}$ vs mode number $\hat{N}$ for e-POD and d-POD modes.\label{fig:convergence_plots_tke}}
\end{figure}

We show the normalized integrated partial TKE vs number of modes in Figure \ref{fig:convergence_plots_tke}a. Visibly faster convergence is obtained using e-POD compared to d-POD. The e-POD modes are ordered by decreasing TKE contribution, producing a concave convergence curve in Figure \ref{fig:convergence_plots_tke}a; however, the d-POD convergence curve is also approximately concave, suggesting that the TKE contribution of each d-POD mode also tends to decrease with mode number, even though no such ordering is explicitly imposed on these modes.

The normalized integrated TKE contribution from a single e-POD or d-POD mode, $\Delta\widetilde{\mathcal{T}}^{\text{e/d},\hat{N}} = \widetilde{\mathcal{T}}^{\text{e/d},\hat{N}} - \widetilde{\mathcal{T}}^{\text{e/d},\hat{N}-1}$, is
\begin{subequations}
\label{eq:tke_modal}
\begin{align}
    \Delta\widetilde{\mathcal{T}}^{\text{e},\hat{N}} &= \frac{\lambda_{\hat{N}}^{\text{e}}}{2\mathcal{T}^{N}}\,,
    \label{eq:tke_epod_modal}
    \\
    \Delta\widetilde{\mathcal{T}}^{\text{d},\hat{N}} &= \frac{\lambda_{\hat{N}}^{\text{d}}}{2\mathcal{T}^{N}} \left\lVert D^{-1} \psi_{\hat{N}}\right\rVert^2_{\mathcal{H}^{\text{e}}}\,.
    \label{eq:tke_dpod_modal}
\end{align}
\end{subequations}

These quantities are shown in in Figure \ref{fig:convergence_plots_tke}b. We note that the plot for $\Delta\widetilde{\mathcal{T}}^{\text{e},\hat{N}}$ is a scaled version of the e-POD spectrum. Compared to e-POD modes, there is a redistribution of TKE contributions from lower to higher d-POD modes, reflecting the slower convergence of the d-POD reconstruction seen in Figure \ref{fig:convergence_plots_tke}a. The d-POD single mode TKE contributions exhibit a decreasing trend across the spectrum, with a more rapid decrease for the first several modes, matching the observed near-concavity of the corresponding plot in Figure \ref{fig:convergence_plots_tke}a. Apart from small fluctuations, the trend in d-POD single mode TKE contributions is in line with the observations made from Figure \ref{fig:all_incremental_profiles}b that the lowest (and thus most dissipative) d-POD modes also tend to be the most TKE-rich.

\subsection{Convergence of integrated dissipation \label{sec:integrated_dissipation_convergence}}
A similar analysis is carried out for the integrated and normalized partial dissipation reconstructions $\widetilde{\mathcal{E}}^{\text{e},\hat{N}}$ obtained from the first $\hat{N}$ e-POD modes using \eqref{eq:epod_mean_dissipation_density}, and $\widetilde{\mathcal{E}}^{\text{e},\hat{N}}$ obtained from the first $\hat{N}$ d-POD modes using \eqref{eq:dpod_mean_dissipation_density},
\begin{align}
    \mathcal{E}^{\text{e/d},\hat{N}} &= \int_{\Omega} \left\langle \varepsilon^{\text{e/d},\hat{N}} \right\rangle \, dx\,,\quad \widetilde{\mathcal{E}}^{\text{e/d},\hat{N}} = \frac{\mathcal{E}^{\text{e/d},\hat{N}}}{\mathcal{E}^{N}}\,,
\end{align}
where $\mathcal{E}^N$ is the full dissipation. The integrated partial dissipations for e-POD and d-POD are, respectively,
\begin{subequations}
\begin{align}
    \mathcal{E}^{\text{e},\hat{N}} &= 2\nu\sum_{n=1}^{\hat{N}} \lambda^{\text{e}}_n \left\lVert D \varphi_n \right\rVert_{\mathcal{H}^{\text{d}}}^2\,,
    \label{eq:dissipation_epod_integrated}
    \\
    \mathcal{E}^{\text{d},\hat{N}} &=  2\nu \sum_{n=1}^{\hat{N}} \lambda^{\text{d}}_n\,.
    \label{eq:dissipation_dpod_integrated}
\end{align}
\end{subequations}

\begin{figure}
    \centerline{\input{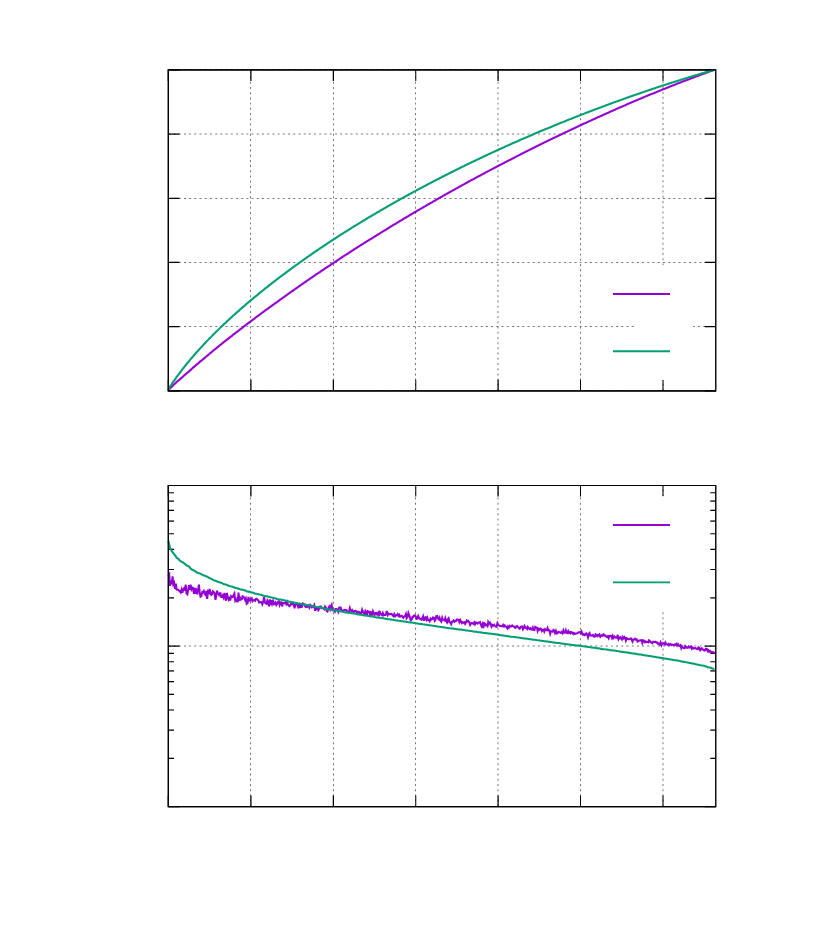}}
    \caption{$(a)$: Reconstruction convergence of integrated dissipation $\widetilde{\mathcal{E}}^{\text{e/d},\hat{N}}$ vs number $\hat{N}$ of e-POD and d-POD modes. $(b)$: Single mode dissipation contribution $\Delta\widetilde{\mathcal{E}}^{\text{e/d},\hat{N}}$ vs mode number $\hat{N}$ for e-POD and d-POD modes. The legend in $(a)$ applies to both plots.}
    \label{fig:convergence_plots_dissipation}
\end{figure}

In Figure \ref{fig:convergence_plots_dissipation}a we show the normalized integrated partial dissipation vs number of modes, illustrating the convergence of the integrated dissipation reconstructions. Both show less rapid convergence than those seen in Figure \ref{fig:convergence_plots_tke}a, implying that either dissipation reconstruction converges more slowly than both TKE reconstructions. The dissipation reconstruction convergence is improved when using d-POD modes compared to e-POD modes, although the improvement is less dramatic than that of TKE convergence when using e-POD over d-POD modes as shown in Figure \ref{fig:convergence_plots_tke}a.

Again, we note that lower modes of either decomposition tend to contribute more to the integrated dissipation, as is also seen from the normalized single-mode dissipation contributions $\Delta \widetilde{\mathcal{E}}^{\text{e/d},\hat{N}}$ shown in Figure \ref{fig:convergence_plots_dissipation}b. These are given by
\begin{subequations}
\label{eq:dissipation_modal}
\begin{align}
    \Delta\widetilde{\mathcal{E}}^{\text{e},\hat{N}} &= \frac{2\nu \lambda_{\hat{N}}^{\text{e}}}{\mathcal{E}^{N}} \left\lVert D \varphi_{\hat{N}}\right\rVert^2_{\mathcal{H}^{\text{d}}}\,,
    \label{eq:dissipation_epod_modal}
    \\
    \Delta\widetilde{\mathcal{E}}^{\text{d},\hat{N}} &= \frac{2\nu \lambda_{\hat{N}}^{\text{d}}}{\mathcal{E}^{N}}\,.
    \label{eq:dissipation_dpod_modal}
\end{align}
\end{subequations}

Analogously to the case of $\Delta\mathcal{T}^{\text{e},\hat{N}}$, the plot of $\Delta\mathcal{E}^{\text{d},\hat{N}}$ in Figure \ref{fig:convergence_plots_dissipation}b is a scaled version of the d-POD spectrum. Both curves in Figure \ref{fig:convergence_plots_dissipation}b appear similar in shape to those in Figure \ref{fig:convergence_plots_tke}b, although they span a narrower range of values, reflecting the modest speed of convergence of both dissipation reconstructions seen in Figure \ref{fig:convergence_plots_dissipation}a. We thus find that for the example considered in this work, using d-POD results in a much more slowly converging reconstruction of integrated TKE compared to e-POD, and only a modest improvement in convergence of the integrated dissipation. 

Convergence curves similar to those seen in Figure \ref{fig:convergence_plots_dissipation} were found by \citet{kostas2005comparison} for enstrophy reconstruction using TKE and enstrophy optimized modes for a backward-facing step flow with $\Rey_h = 580$, where $\Rey_h$ is the Reynolds number derived from bulk velocity and step height. The comparison between enstrophy and dissipation is justified by both being second order gradient derived quantities, which are proportional in the homogeneous case which was found by \citet{tardu2017near} to be an excellent approximation across the width of the channel. \citet{kostas2005comparison} also considered a high Reynolds number case ($\Rey_h = 4680$) which displayed a more pronounced efficiency advantage when reconstructing enstrophy using enstrophy-optimized over TKE-optimized modes. In both the high and low Reynolds number case, lower e-POD modes tended to contribute more enstrophy.

Comparing the energetic large-scale structures and dissipative small-scale structures, the greater number of degrees of freedom available to the latter makes an efficient low-dimensional representation difficult: while one may conceive of the largest scale of a flow being dominated by one or few "big eddies", there will typically be a much larger number of "small eddies" on the small scale. 

While the lowest modes of either decomposition are by construction those with the largest contributions to the quantity with respect to which the decomposition is optimized (TKE for e-POD modes and dissipation for d-POD modes), we find that this also tends to be the case with respect to the quantity with respect to which it is \textit{not} optimized. This suggests that the lowest modes in either decomposition capture structures that are both coherent across scales and active for appreciable fractions of the simulation time. On the other hand, the structures have limited spatial coherence, as shown by the drop in efficiency of the d-POD TKE reconstruction when moving into the bulk.

\subsection{Eigenvalue-normalized single-mode contributions\label{sec:ensmc}}
Formally, the fact that the lowest modes in each basis appear to contribute the most to each reconstruction is linked to the mean modal coefficient entering via eigenvalues in the reconstructions in \eqref{eq:epod_mean_tke_density}--\eqref{eq:dpod_mean_dissipation_density}. Here we investigate the contributions inherent to the structures encoded by modes by separating them from the eigenvalue.

\begin{figure}
    \centerline{\input{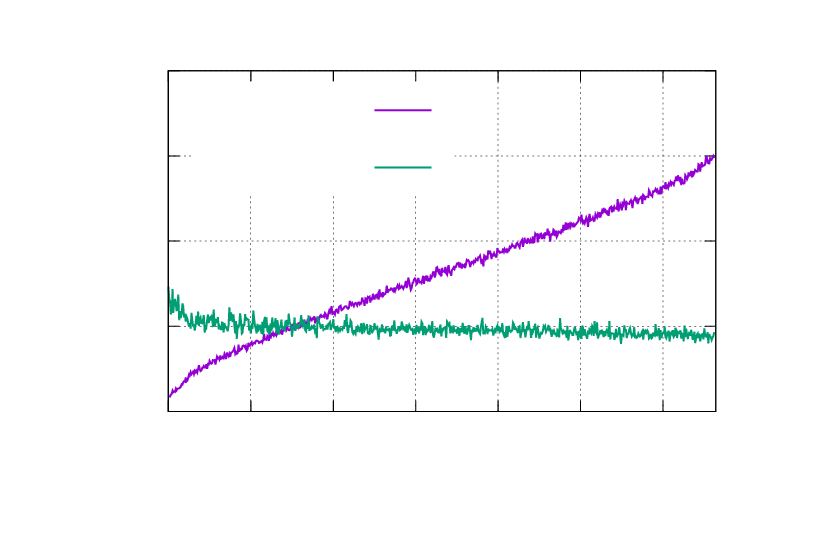}}
    \caption{Eigenvalue-normalized single mode contributions (ENSMCs) for e-POD dissipation reconstruction and d-POD TKE reconstruction, plotted vs mode number $\hat{N}$.
    \label{fig:ev_normalized_mc}}
\end{figure}

The eigenvalue associated with each mode is the mean square modal coefficient (cf.\  \eqref{eq:coefficent_correlation}) and governs the contribution from the mode to second order quantities like TKE and dissipation (cf.\ \eqref{eq:tke_modal} and \eqref{eq:dissipation_modal}). The decreasing single mode contributions to either reconstructed quantity using the optimal bases, $\Delta\widetilde{\mathcal{T}}^{\text{e},\hat{N}}$ and $\Delta\widetilde{\mathcal{E}}^{\text{d},\hat{N}}$, thus follows directly from the ordering of modes by decreasing eigenvalue. The observations in the preceding two sections of similar trends for single mode contributions using non-optimal bases, $\Delta\widetilde{\mathcal{T}}^{\text{d},\hat{N}}$ and $\Delta\widetilde{\mathcal{E}}^{\text{e},\hat{N}}$, can be linked to the scaling in \eqref{eq:tke_dpod_modal} and \eqref{eq:dissipation_epod_modal} with the corresponding eigenvalues, though this eigenvalue dependence is modulated by the mode-derived squared norms $\left\lVert D^{-1}\psi_n\right\rVert^2_{\mathcal{H}^{\text{e}}}$ and $\left\lVert D \varphi_n\right\rVert^2_{\mathcal{H}^{\text{d}}}$. These latter two factors may be considered as measures of, respectively, the inherent TKE content of the structures encoded in d-POD modes, and the inherent dissipation content of the structures encoded in e-POD modes. The eigenvalue factor accounts for the time-dependent amplitude with which the modes enter the reconstructions. To isolate the effect of the mode-derived squared norms, we consider the \emph{eigenvalue-normalized single mode contributions} (ENSMCs) which are proportional to the field square norms,
\begin{subequations}
\begin{align}
    \Delta\widetilde{\mathcal{E}}^{\text{e},\hat{N}}/\widetilde{\lambda}^\text{e}_{\hat{N}} &\propto \left\lVert D\varphi_{\hat{N}}\right\rVert^2_{\mathcal{H}^{\text{d}}}\,,
    \label{eq:epod_disp_enmc}
    \\
    \Delta\widetilde{\mathcal{T}}^{\text{d},\hat{N}}/\widetilde{\lambda}^{\text{d}}_{\hat{N}} & \propto \left\lVert D^{-1}\psi_{\hat{N}}\right\rVert^2_{\mathcal{H}^{\text{e}}}\,,
    \label{eq:dpod_tke_enmc}
\end{align}
\end{subequations}
where $\widetilde{\lambda}_{\hat{N}}^{\text{e}}$ and $\widetilde{\lambda}_{\hat{N}}^{\text{d}}$ are the $\hat{N}$\textsuperscript{th} e-POD or d-POD eigenvalue, respectively, normalized by the sum of all eigenvalues in the corresponding decomposition. The ENSMCs are plotted against mode number $\hat{N}$ in Figure \ref{fig:ev_normalized_mc}.

The e-POD dissipation ENSMCs \eqref{eq:epod_disp_enmc} follow an increasing trend with mode number. Higher e-POD modes, representing lower energy, thus tend to contribute more to dissipation reconstruction \emph{relative to their mean square modal coefficient in the reconstruction}, implying that the structures encoded in such modes are more strongly \emph{inherently} dissipative - although it should be emphasized that due to the eigenvalue factors the actual contribution to dissipation exhibits the opposite trend.

A rather different behavior is found for the d-POD TKE reconstruction ENSMCs \eqref{eq:dpod_tke_enmc}. Apart from a decrease for the first several modes followed by a slightly decreasing trend for the remaining modes, overall we find these to be nearly constant with fluctuations around unity. This implies that all d-POD modes describe structures of roughly similar inherent TKE content, with the different single mode TKE contributions almost entirely due to varying modal coefficients encoded by eigenvalues.

The different trends exhibited by the ENSMCs suggest that when considering inherent contributions, TKE-rich structures encoded in low e-POD modes tend to be weakly dissipative, and those poor in TKE, encoded in high e-POD modes, are more strongly dissipative. At the same time, however, weakly and strongly dissipative structures encoded by d-POD modes are all nearly equally rich in TKE. TKE is carried primarily by large-scale structures, which are therefore encoded by lower e-POD modes, leaving any remaining small-scale structure to be resolved by higher modes. This small-scale structure is associated with larger gradients, leading to greater inherent dissipation content in higher e-POD modes.

\begin{figure*}[t]
    \centerline{\input{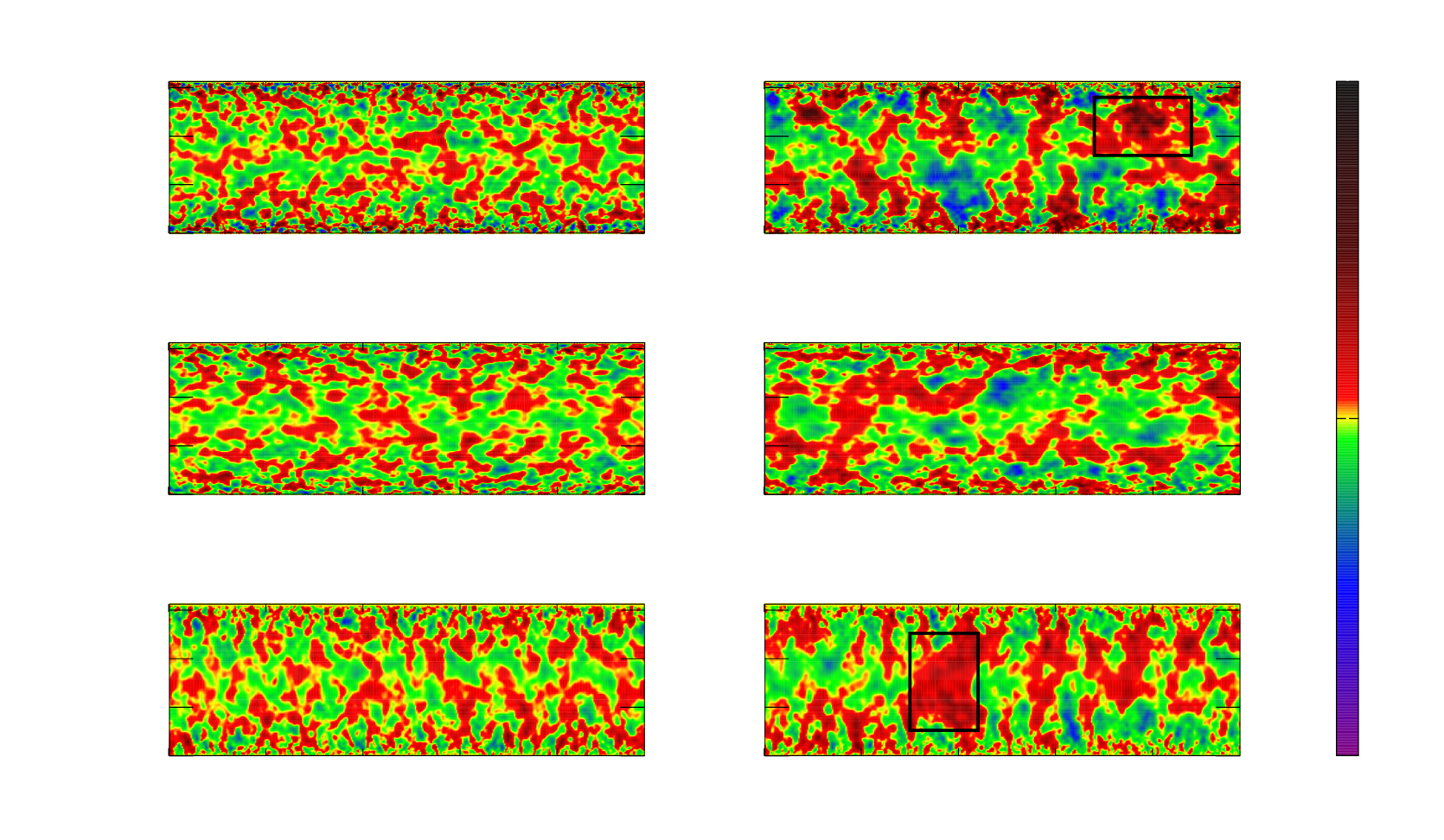}}
    \caption{\emph{Left:} Components $\varphi^i_{500}$ of the 500\textsuperscript{th} e-POD mode. \emph{Right:} Components $(\widetilde{D^{-1}\psi}_{500})^i$ of the normalized velocity field corresponding to the 500\textsuperscript{th} d-POD mode. Boxes in $(\widetilde{D^{-1}\psi}_{500})^1$ and $(\widetilde{D^{-1}\psi}_{500})^3$ mark examples of large-scale structures identified in this mode. Axes show coordinates in wall units (spanwise horizontal, transverse vertical), and the color scale shows values with the full range normalized to $[-1;1]$.\label{fig:high_mode}}
\end{figure*}

Conversely, as suggested by the mode plots considered in Section \ref{sec:results_modes}, low d-POD modes encode structures which cover a wide range of scales. As all features of the flow may eventually be resolved in terms of such structures, we hypothesize that a wide spectral range may be characteristic for d-POD modes throughout the spectrum. If this is the case then the decomposition does not leave a residual of large-scale TKE-rich structures to be resolved by higher modes; instead, the TKE ENSMC remains stable across d-POD modes, as we observe. This hypothesis may also explain the modest range of less than one decade of d-POD single mode dissipation contributions, contrasted with the range of nearly two decades of e-POD single mode TKE contributions shown in Figures \ref{fig:convergence_plots_tke}b and \ref{fig:convergence_plots_dissipation}b.

A thorough test of this hypothesis could be carried out e.g.\ by computing spanwise spectra of e-POD modes and d-POD velocity fields, and analyze how the width and peaks of such spectra vary as a function of mode number. This is in principle equivalent to the SPOD analysis discussed at the end of Section \ref{sec:channel_flow}. While such an analysis is beyond the scope of this work, we gain an indication of the viability of the hypothesis by plotting one of the higher e-POD modes, $\varphi_{500}$, along with the corresponding d-POD velocity field, $D^{-1}\psi_{500}$, in Figure \ref{fig:high_mode}. While large-scale structures cannot be readily identified in $\varphi_{500}$, we find large-scale structures clearly exhibited in the bulk in $D^{-1}\psi_{500}$ (examples marked by boxes in Figure \ref{fig:high_mode}). This lends support to the hypothesis laid out above, and indicates that while the constraint of completeness of the basis forces e-POD modes to adhere to a hierarchy of scales, causing higher modes to describe TKE-poor structures, the nature of dissipation does not lead to a similar inverted effect for d-POD modes.

\section{Discussion\label{sec:discussion}}
Two important differences between the e-POD and the d-POD can be identified based on the observations presented in the preceding sections. 

Firstly, since TKE and dissipation have different spatial distributions in the channel flow profile, the decompositions emphasize structures in the regions where the respective quantities dominate. While e-POD prioritizes structures in the bulk, d-POD prioritizes those near the wall, and either decomposition reconstructs both TKE and dissipation most efficiently in the region prioritized by that decomposition. We therefore hypothesize that modeling of TKE-poor but dynamically important regions such as the near-wall region in the channel flow may benefit from incorporating d-POD modes into the model basis.

Secondly, combining the analysis of \mbox{ENSMCs} with the observed features in low and high e-POD modes and d-POD velocity fields supports the assignment of different characteristic scales to TKE-rich and dissipative structures. As also suggested by spectral theory the lowest e-POD modes encode mainly large-scale structures within the TKE-rich bulk, and smaller scale structures in the near-wall region, consigning smaller scale bulk structures to higher modes. However, the d-POD does not represent a simple inversion of these scales, as one might perhaps expect, in which low modes represent the small dissipative scales and higher modes larger energetic scales. We find instead that d-POD modes represent a wide range of scales throughout their spectrum. The contributions of d-POD modes in the reconstruction of TKE are therefore determined mainly by the time-dependent amplitude of each mode, as the modes themselves exhibit approximately constant inherent TKE content.

The analysis presented here is based on $\num{\thensamples}$ snapshots separated by $T < T_E$, where $T_E$ is the integral time scale obtained from the temporal correlation function in the center of the channel. The moderate separation indicates that samples are not fully independent, and combined with the limited number of realizations it is probable that the full dynamics of the flow have not been captured. This does not invalidate the analysis made in this work, although it may cause additional coupling between e-POD and d-POD modes that would not have appeared with a more complete sampling. Such a coupling would lessen the observed advantages of the d-POD over the e-POD with regards to representation of near-wall structures and the rate of convergence of dissipation. A larger sample set and more independent sampling could e.g.\ be achieved using measured rather than simulated data, although this presents a different set of challenges, in particular with regards to resolving the relevant scales. However, given that the sampling problem can be resolved we would expect the resulting d-POD dissipation reconstruction to have a greater convergence lead over e-POD, as well as significant benefits when building flow models using an e-POD basis augmented with d-POD modes.

\section{Conclusions\label{sec:conclusion}}
The dissipation-optimized proper orthogonal decomposition (d-POD) is introduced as a supplement to the conventional TKE-optimized proper orthogonal decomposition (e-POD). A spectral inverse strain rate tensor (SRT) operator is presented, which maps d-POD SRT modes to the underlying velocity fields. The formulation of the inverse spectral SRT operator may be straightforwardly generalized to other maps used for decompositions based on snapshot-derived quantities. Combining the d-POD and the spectral inverse SRT operator allows for dissipation-optimized expansion of flow fields and reconstruction of any flow-dependent quantity.

This method is applied along with e-POD to a data set produced by direct numerical simulation of a turbulent periodic channel flow \mbox{($\Rey_{\tau}=\renum$)}. The resulting modes, combined with analysis of single mode contributions to reconstruction of TKE and dissipation, suggest that the spatial scales of structures encoded in e-POD modes tend to decrease with increasing mode number, whereas structures encoded by d-POD modes span a wider and approximately constant range of spatial scales throughout the range of mode numbers.

Transverse profiles of TKE and dissipation are reconstructed using both e-POD and d-POD. The spatial distribution of TKE and dissipation across the channel profile is found to be crucial in determining the relative reconstruction efficiency of the two decompositions, with both TKE and dissipation reconstructed more efficiently in the dissipation-rich near-wall region using d-POD, and in the TKE-rich bulk using e-POD. 

The d-POD basis thus complements the e-POD basis by prioritizing scales or regions of the flow that are given low priority in the e-POD, efficiently representing structures that would require a larger number of e-POD modes to represent accurately. Combining the two decompositions in reduced order models may prove useful for including otherwise unresolved flow structures, potentially enhancing model efficiency. In particular, dissipation is a key parameter in turbulence models and theory, which is in general not well captured by e-POD modes. Since the two decompositions are formed independently of each other, it would be relevant in future studies to address the combined efficiency with which structures are represented in such a basis.

\section*{Acknowledgments}
P.J.O. acknowledges financial support from the Poul Due Jensen Foundation: Financial support from the Poul Due Jensen Foundation (Grundfos Foundation) for this research is gratefully acknowledged.

A.H. and C.M.V. acknowledge financial support from the European Research Council: This project has received funding from the European Research Council (ERC) under the European Union's Horizon 2020 research and innovation programme (grant agreement No 803419). 

The authors gratefully acknowledge the computational and data resources provided on the Sophia HPC Cluster at the Technical University of Denmark, DOI: 10.57940/FAFC-6M81.

\section*{Author declarations}
\subsection*{Conflict of interest}
The authors have no conflicts to disclose.

\subsection*{Author contributions}
\mbox{\textbf{Peder J.\ Olesen:}} Conceptualization (equal); Data curation (equal); Formal analysis (equal); Methodology (equal); Project administration (equal); Software (equal); Validation (equal); Visualization (lead); Writing -- original draft preparation (lead); Writing -- review and editing (lead).
\mbox{\textbf{Azur Hod\v{z}i\'{c}:}} Conceptualization (equal); Formal analysis (equal); Methodology (equal); Supervision (equal); Writing -- review and editing (equal).
\mbox{\textbf{Søren J.\ Andersen:}} Data curation (lead); Formal analysis (lead); Investigation (lead); Methodology (equal); Project administration (equal); Resources (equal); Software (equal); Validation (equal); Writing -- original draft preparation (support); Writing -- review and editing (support).
\mbox{\textbf{Niels N.\ Sørensen:}} Data curation (equal); Formal analysis (equal); Investigation (equal); Methodology (equal); Resources (equal); Software (equal); Validation (equal).
\mbox{\textbf{Clara M.\ Velte:}} Conceptualization (equal); Funding acquisition (lead); Methodology (equal); Project administration (equal); Resources (equal); Supervision (equal); Writing -- review and editing (equal).

\section*{Data availability}
Raw data were generated at the Sophia HPC Cluster at the Technical University of Denmark large scale facility.  Derived data supporting the findings of this study are available from the corresponding author upon reasonable request.

\bibliography{references.bib}

%aipnauth4-2.bst 2018-12-27 (MD) hand-edited version of apsauth4-1.bst
%Control: author (9) reversed initials
\begin{thebibliography}{39}%
\makeatletter
\providecommand \@ifxundefined [1]{%
 \@ifx{#1\undefined}
}%
\providecommand \@ifnum [1]{%
 \ifnum #1\expandafter \@firstoftwo
 \else \expandafter \@secondoftwo
 \fi
}%
\providecommand \@ifx [1]{%
 \ifx #1\expandafter \@firstoftwo
 \else \expandafter \@secondoftwo
 \fi
}%
\providecommand \natexlab [1]{#1}%
\providecommand \enquote  [1]{``#1''}%
\providecommand \bibnamefont  [1]{#1}%
\providecommand \bibfnamefont [1]{#1}%
\providecommand \citenamefont [1]{#1}%
\providecommand \href@noop [0]{\@secondoftwo}%
\providecommand \href [0]{\begingroup \@sanitize@url \@href}%
\providecommand \@href[1]{\@@startlink{#1}\@@href}%
\providecommand \@@href[1]{\endgroup#1\@@endlink}%
\providecommand \@sanitize@url [0]{\catcode `\\12\catcode `\$12\catcode
  `\&12\catcode `\#12\catcode `\^12\catcode `\_12\catcode `\%12\relax}%
\providecommand \@@startlink[1]{}%
\providecommand \@@endlink[0]{}%
\providecommand \url  [0]{\begingroup\@sanitize@url \@url }%
\providecommand \@url [1]{\endgroup\@href {#1}{\urlprefix }}%
\providecommand \urlprefix  [0]{URL }%
\providecommand \Eprint [0]{\href }%
\providecommand \doibase [0]{https://doi.org/}%
\providecommand \selectlanguage [0]{\@gobble}%
\providecommand \bibinfo  [0]{\@secondoftwo}%
\providecommand \bibfield  [0]{\@secondoftwo}%
\providecommand \translation [1]{[#1]}%
\providecommand \BibitemOpen [0]{}%
\providecommand \bibitemStop [0]{}%
\providecommand \bibitemNoStop [0]{.\EOS\space}%
\providecommand \EOS [0]{\spacefactor3000\relax}%
\providecommand \BibitemShut  [1]{\csname bibitem#1\endcsname}%
\let\auto@bib@innerbib\@empty
%</preamble>
\bibitem [{\citenamefont {Ali}, \citenamefont {Kadum},\ and\ \citenamefont
  {Cal}(2016)}]{ali2016focused}%
  \BibitemOpen
  \bibfield  {author} {\bibinfo {author} {\bibnamefont {Ali}, \bibfnamefont
  {N.}}, \bibinfo {author} {\bibnamefont {Kadum}, \bibfnamefont {H.~F.}}, and\
  \bibinfo {author} {\bibnamefont {Cal}, \bibfnamefont {R.~B.}},\ }\bibfield
  {title} {\enquote {\bibinfo {title} {Focused-based multifractal analysis of
  the wake in a wind turbine array utilizing proper orthogonal
  decomposition},}\ }\href@noop {} {\bibfield  {journal} {\bibinfo  {journal}
  {Journal of Renewable and Sustainable Energy}\ }\textbf {\bibinfo {volume}
  {8}},\ \bibinfo {pages} {063306} (\bibinfo {year} {2016})}\BibitemShut
  {NoStop}%
\bibitem [{\citenamefont {Andersen}\ and\ \citenamefont
  {Murcia~Leon}(2022)}]{PS-ROM2022}%
  \BibitemOpen
  \bibfield  {author} {\bibinfo {author} {\bibnamefont {Andersen},
  \bibfnamefont {S.~J.}}and\ \bibinfo {author} {\bibnamefont {Murcia~Leon},
  \bibfnamefont {J.~P.}},\ }\bibfield  {title} {\enquote {\bibinfo {title}
  {Predictive and stochastic reduced order modelling of wind turbine wake
  dynamics},}\ }\href {https://doi.org/10.5194/wes-2022-45} {\bibfield
  {journal} {\bibinfo  {journal} {Wind Energy Science Discussions}\ }\textbf
  {\bibinfo {volume} {2022}},\ \bibinfo {pages} {1--27} (\bibinfo {year}
  {2022})}\BibitemShut {NoStop}%
\bibitem [{\citenamefont {Aubry}\ \emph {et~al.}(1988)\citenamefont {Aubry},
  \citenamefont {Holmes}, \citenamefont {Lumley},\ and\ \citenamefont
  {Stone}}]{aubry1988dynamics}%
  \BibitemOpen
  \bibfield  {author} {\bibinfo {author} {\bibnamefont {Aubry}, \bibfnamefont
  {N.}}, \bibinfo {author} {\bibnamefont {Holmes}, \bibfnamefont {P.}},
  \bibinfo {author} {\bibnamefont {Lumley}, \bibfnamefont {J.~L.}}, and\
  \bibinfo {author} {\bibnamefont {Stone}, \bibfnamefont {E.}},\ }\bibfield
  {title} {\enquote {\bibinfo {title} {The dynamics of coherent structures in
  the wall region of a turbulent boundary layer},}\ }\href@noop {} {\bibfield
  {journal} {\bibinfo  {journal} {Journal of Fluid Mechanics}\ }\textbf
  {\bibinfo {volume} {192}},\ \bibinfo {pages} {115--173} (\bibinfo {year}
  {1988})}\BibitemShut {NoStop}%
\bibitem [{\citenamefont {Balay}\ \emph {et~al.}(2021)\citenamefont {Balay},
  \citenamefont {Abhyankar}, \citenamefont {Adams}, \citenamefont {Benson},
  \citenamefont {Brown}, \citenamefont {Brune}, \citenamefont {Buschelman},
  \citenamefont {Constantinescu}, \citenamefont {Dalcin}, \citenamefont
  {Dener}, \citenamefont {Eijkhout}, \citenamefont {Gropp}, \citenamefont
  {Hapla}, \citenamefont {Isaac}, \citenamefont {Jolivet}, \citenamefont
  {Karpeev}, \citenamefont {Kaushik}, \citenamefont {Knepley}, \citenamefont
  {Kong}, \citenamefont {Kruger}, \citenamefont {May}, \citenamefont {McInnes},
  \citenamefont {Mills}, \citenamefont {Mitchell}, \citenamefont {Munson},
  \citenamefont {Roman}, \citenamefont {Rupp}, \citenamefont {Sanan},
  \citenamefont {Sarich}, \citenamefont {Smith}, \citenamefont {Zampini},
  \citenamefont {Zhang}, \citenamefont {Zhang},\ and\ \citenamefont
  {Zhang}}]{petsc-user-ref}%
  \BibitemOpen
  \bibfield  {author} {\bibinfo {author} {\bibnamefont {Balay}, \bibfnamefont
  {S.}}, \bibinfo {author} {\bibnamefont {Abhyankar}, \bibfnamefont {S.}},
  \bibinfo {author} {\bibnamefont {Adams}, \bibfnamefont {M.~F.}}, \bibinfo
  {author} {\bibnamefont {Benson}, \bibfnamefont {S.}}, \bibinfo {author}
  {\bibnamefont {Brown}, \bibfnamefont {J.}}, \bibinfo {author} {\bibnamefont
  {Brune}, \bibfnamefont {P.}}, \bibinfo {author} {\bibnamefont {Buschelman},
  \bibfnamefont {K.}}, \bibinfo {author} {\bibnamefont {Constantinescu},
  \bibfnamefont {E.~M.}}, \bibinfo {author} {\bibnamefont {Dalcin},
  \bibfnamefont {L.}}, \bibinfo {author} {\bibnamefont {Dener}, \bibfnamefont
  {A.}}, \bibinfo {author} {\bibnamefont {Eijkhout}, \bibfnamefont {V.}},
  \bibinfo {author} {\bibnamefont {Gropp}, \bibfnamefont {W.~D.}}, \bibinfo
  {author} {\bibnamefont {Hapla}, \bibfnamefont {V.}}, \bibinfo {author}
  {\bibnamefont {Isaac}, \bibfnamefont {T.}}, \bibinfo {author} {\bibnamefont
  {Jolivet}, \bibfnamefont {P.}}, \bibinfo {author} {\bibnamefont {Karpeev},
  \bibfnamefont {D.}}, \bibinfo {author} {\bibnamefont {Kaushik}, \bibfnamefont
  {D.}}, \bibinfo {author} {\bibnamefont {Knepley}, \bibfnamefont {M.~D.}},
  \bibinfo {author} {\bibnamefont {Kong}, \bibfnamefont {F.}}, \bibinfo
  {author} {\bibnamefont {Kruger}, \bibfnamefont {S.}}, \bibinfo {author}
  {\bibnamefont {May}, \bibfnamefont {D.~A.}}, \bibinfo {author} {\bibnamefont
  {McInnes}, \bibfnamefont {L.~C.}}, \bibinfo {author} {\bibnamefont {Mills},
  \bibfnamefont {R.~T.}}, \bibinfo {author} {\bibnamefont {Mitchell},
  \bibfnamefont {L.}}, \bibinfo {author} {\bibnamefont {Munson}, \bibfnamefont
  {T.}}, \bibinfo {author} {\bibnamefont {Roman}, \bibfnamefont {J.~E.}},
  \bibinfo {author} {\bibnamefont {Rupp}, \bibfnamefont {K.}}, \bibinfo
  {author} {\bibnamefont {Sanan}, \bibfnamefont {P.}}, \bibinfo {author}
  {\bibnamefont {Sarich}, \bibfnamefont {J.}}, \bibinfo {author} {\bibnamefont
  {Smith}, \bibfnamefont {B.~F.}}, \bibinfo {author} {\bibnamefont {Zampini},
  \bibfnamefont {S.}}, \bibinfo {author} {\bibnamefont {Zhang}, \bibfnamefont
  {H.}}, \bibinfo {author} {\bibnamefont {Zhang}, \bibfnamefont {H.}}, and\
  \bibinfo {author} {\bibnamefont {Zhang}, \bibfnamefont {J.}},\ }\href@noop {}
  {\enquote {\bibinfo {title} {{PETSc/TAO} users manual},}\ }\bibinfo {type}
  {Tech. Rep.}\ \bibinfo {number} {ANL-21/39 - Revision 3.16}\ (\bibinfo
  {institution} {Argonne National Laboratory},\ \bibinfo {year}
  {2021})\BibitemShut {NoStop}%
\bibitem [{\citenamefont {Balay}\ \emph {et~al.}(1997)\citenamefont {Balay},
  \citenamefont {Gropp}, \citenamefont {McInnes},\ and\ \citenamefont
  {Smith}}]{Balay1997}%
  \BibitemOpen
  \bibfield  {author} {\bibinfo {author} {\bibnamefont {Balay}, \bibfnamefont
  {S.}}, \bibinfo {author} {\bibnamefont {Gropp}, \bibfnamefont {W.~D.}},
  \bibinfo {author} {\bibnamefont {McInnes}, \bibfnamefont {L.~C.}}, and\
  \bibinfo {author} {\bibnamefont {Smith}, \bibfnamefont {B.~F.}},\ }\enquote
  {\bibinfo {title} {Efficient management of parallelism in object-oriented
  numerical software libraries},}\ in\ \href
  {https://doi.org/10.1007/978-1-4612-1986-6_8} {\emph {\bibinfo {booktitle}
  {Modern Software Tools for Scientific Computing}}},\ \bibinfo {editor}
  {edited by\ \bibinfo {editor} {\bibfnamefont {E.}~\bibnamefont {Arge}},
  \bibinfo {editor} {\bibfnamefont {A.~M.}\ \bibnamefont {Bruaset}}, \ and\
  \bibinfo {editor} {\bibfnamefont {H.~P.}\ \bibnamefont {Langtangen}}}\
  (\bibinfo  {publisher} {Birkh{\"a}user Boston},\ \bibinfo {address} {Boston,
  MA},\ \bibinfo {year} {1997})\ pp.\ \bibinfo {pages} {163--202}\BibitemShut
  {NoStop}%
\bibitem [{\citenamefont {Bergmann}, \citenamefont {Bruneau},\ and\
  \citenamefont {Iollo}(2009)}]{bergmann2009enablers}%
  \BibitemOpen
  \bibfield  {author} {\bibinfo {author} {\bibnamefont {Bergmann},
  \bibfnamefont {M.}}, \bibinfo {author} {\bibnamefont {Bruneau}, \bibfnamefont
  {C.-H.}}, and\ \bibinfo {author} {\bibnamefont {Iollo}, \bibfnamefont {A.}},\
  }\bibfield  {title} {\enquote {\bibinfo {title} {Enablers for robust {POD}
  models},}\ }\href@noop {} {\bibfield  {journal} {\bibinfo  {journal} {Journal
  of Computational Physics}\ }\textbf {\bibinfo {volume} {228}},\ \bibinfo
  {pages} {516--538} (\bibinfo {year} {2009})}\BibitemShut {NoStop}%
\bibitem [{\citenamefont {Berkooz}, \citenamefont {Holmes},\ and\ \citenamefont
  {Lumley}(1993)}]{berkooz1993proper}%
  \BibitemOpen
  \bibfield  {author} {\bibinfo {author} {\bibnamefont {Berkooz}, \bibfnamefont
  {G.}}, \bibinfo {author} {\bibnamefont {Holmes}, \bibfnamefont {P.}}, and\
  \bibinfo {author} {\bibnamefont {Lumley}, \bibfnamefont {J.~L.}},\ }\bibfield
   {title} {\enquote {\bibinfo {title} {The proper orthogonal decomposition in
  the analysis of turbulent flows},}\ }\href@noop {} {\bibfield  {journal}
  {\bibinfo  {journal} {Annual Review of Fluid Mechanics}\ }\textbf {\bibinfo
  {volume} {25}},\ \bibinfo {pages} {539--575} (\bibinfo {year}
  {1993})}\BibitemShut {NoStop}%
\bibitem [{\citenamefont {Bermejo-Moreno}, \citenamefont {Pullin},\ and\
  \citenamefont {Horiuti}(2009)}]{bermejo2009geometry}%
  \BibitemOpen
  \bibfield  {author} {\bibinfo {author} {\bibnamefont {Bermejo-Moreno},
  \bibfnamefont {I.}}, \bibinfo {author} {\bibnamefont {Pullin}, \bibfnamefont
  {D.~I.}}, and\ \bibinfo {author} {\bibnamefont {Horiuti}, \bibfnamefont
  {K.}},\ }\bibfield  {title} {\enquote {\bibinfo {title} {Geometry of
  enstrophy and dissipation, grid resolution effects and proximity issues in
  turbulence},}\ }\href@noop {} {\bibfield  {journal} {\bibinfo  {journal}
  {Journal of Fluid Mechanics}\ }\textbf {\bibinfo {volume} {620}},\ \bibinfo
  {pages} {121--166} (\bibinfo {year} {2009})}\BibitemShut {NoStop}%
\bibitem [{\citenamefont {Citriniti}\ and\ \citenamefont
  {George}(2000)}]{citriniti2000reconstruction}%
  \BibitemOpen
  \bibfield  {author} {\bibinfo {author} {\bibnamefont {Citriniti},
  \bibfnamefont {J.~H.}}and\ \bibinfo {author} {\bibnamefont {George},
  \bibfnamefont {W.~K.}},\ }\bibfield  {title} {\enquote {\bibinfo {title}
  {Reconstruction of the global velocity field in the axisymmetric mixing layer
  utilizing the proper orthogonal decomposition},}\ }\href@noop {} {\bibfield
  {journal} {\bibinfo  {journal} {Journal of Fluid Mechanics}\ }\textbf
  {\bibinfo {volume} {418}},\ \bibinfo {pages} {137--166} (\bibinfo {year}
  {2000})}\BibitemShut {NoStop}%
\bibitem [{\citenamefont {Couplet}, \citenamefont {Sagaut},\ and\ \citenamefont
  {Basdevant}(2003)}]{couplet2003intermodal}%
  \BibitemOpen
  \bibfield  {author} {\bibinfo {author} {\bibnamefont {Couplet}, \bibfnamefont
  {M.}}, \bibinfo {author} {\bibnamefont {Sagaut}, \bibfnamefont {P.}}, and\
  \bibinfo {author} {\bibnamefont {Basdevant}, \bibfnamefont {C.}},\ }\bibfield
   {title} {\enquote {\bibinfo {title} {Intermodal energy transfers in a proper
  orthogonal decomposition--galerkin representation of a turbulent separated
  flow},}\ }\href@noop {} {\bibfield  {journal} {\bibinfo  {journal} {Journal
  of Fluid Mechanics}\ }\textbf {\bibinfo {volume} {491}},\ \bibinfo {pages}
  {275--284} (\bibinfo {year} {2003})}\BibitemShut {NoStop}%
\bibitem [{\citenamefont {Fahl}(2001)}]{fahl2001computation}%
  \BibitemOpen
  \bibfield  {author} {\bibinfo {author} {\bibnamefont {Fahl}, \bibfnamefont
  {M.}},\ }\bibfield  {title} {\enquote {\bibinfo {title} {Computation of {POD}
  basis functions for fluid flows with {L}ánczos methods},}\ }\href@noop {}
  {\bibfield  {journal} {\bibinfo  {journal} {Mathematical and Computer
  Modelling}\ }\textbf {\bibinfo {volume} {34}},\ \bibinfo {pages} {91--107}
  (\bibinfo {year} {2001})}\BibitemShut {NoStop}%
\bibitem [{\citenamefont {Gatski}\ and\ \citenamefont
  {Glauser}(1992)}]{Gatski1992}%
  \BibitemOpen
  \bibfield  {author} {\bibinfo {author} {\bibnamefont {Gatski}, \bibfnamefont
  {T.~B.}}and\ \bibinfo {author} {\bibnamefont {Glauser}, \bibfnamefont
  {M.~N.}},\ }\bibfield  {title} {\enquote {\bibinfo {title} {Proper orthogonal
  decomposition based turbulence modeling},}\ }in\ \href@noop {} {\emph
  {\bibinfo {booktitle} {Instability, Transition, and Turbulence}}},\ \bibinfo
  {editor} {edited by\ \bibinfo {editor} {\bibfnamefont {M.~Y.}\ \bibnamefont
  {Hussaini}}, \bibinfo {editor} {\bibfnamefont {A.}~\bibnamefont {Kumar}}, \
  and\ \bibinfo {editor} {\bibfnamefont {C.~L.}\ \bibnamefont {Streett}}}\
  (\bibinfo  {publisher} {Springer New York},\ \bibinfo {address} {New York,
  NY},\ \bibinfo {year} {1992})\ pp.\ \bibinfo {pages} {498--510}\BibitemShut
  {NoStop}%
\bibitem [{\citenamefont {Grimberg}, \citenamefont {Farhat},\ and\
  \citenamefont {Youkilis}(2020)}]{grimberg2020stability}%
  \BibitemOpen
  \bibfield  {author} {\bibinfo {author} {\bibnamefont {Grimberg},
  \bibfnamefont {S.}}, \bibinfo {author} {\bibnamefont {Farhat}, \bibfnamefont
  {C.}}, and\ \bibinfo {author} {\bibnamefont {Youkilis}, \bibfnamefont {N.}},\
  }\bibfield  {title} {\enquote {\bibinfo {title} {On the stability of
  projection-based model order reduction for convection-dominated laminar and
  turbulent flows},}\ }\href@noop {} {\bibfield  {journal} {\bibinfo  {journal}
  {Journal of Computational Physics}\ }\textbf {\bibinfo {volume} {419}},\
  \bibinfo {pages} {109681} (\bibinfo {year} {2020})}\BibitemShut {NoStop}%
\bibitem [{\citenamefont {Holmes}\ \emph {et~al.}(2012)\citenamefont {Holmes},
  \citenamefont {Lumley}, \citenamefont {Berkooz},\ and\ \citenamefont
  {Rowley}}]{holmes2012turbulence}%
  \BibitemOpen
  \bibfield  {author} {\bibinfo {author} {\bibnamefont {Holmes}, \bibfnamefont
  {P.}}, \bibinfo {author} {\bibnamefont {Lumley}, \bibfnamefont {J.~L.}},
  \bibinfo {author} {\bibnamefont {Berkooz}, \bibfnamefont {G.}}, and\ \bibinfo
  {author} {\bibnamefont {Rowley}, \bibfnamefont {C.~W.}},\ }\href@noop {}
  {\emph {\bibinfo {title} {Turbulence, Coherent Structures, Dynamical Systems
  and Symmetry}}}\ (\bibinfo  {publisher} {Cambridge university press},\
  \bibinfo {year} {2012})\BibitemShut {NoStop}%
\bibitem [{\citenamefont {Huang}(1994)}]{huang1994limitations}%
  \BibitemOpen
  \bibfield  {author} {\bibinfo {author} {\bibnamefont {Huang}, \bibfnamefont
  {H.}},\ }\emph {\bibinfo {title} {Limitations of and improvements to {PIV}
  and its application to a backward-facing step flow}},\ \href@noop {}
  {\bibinfo {type} {{PhD} thesis}},\ \bibinfo  {school} {Technische
  Universität Berlin} (\bibinfo {year} {1994})\BibitemShut {NoStop}%
\bibitem [{\citenamefont {Iollo}\ \emph {et~al.}(2000)\citenamefont {Iollo},
  \citenamefont {Dervieux}, \citenamefont {D{\'e}sid{\'e}ri},\ and\
  \citenamefont {Lanteri}}]{iollo2000two}%
  \BibitemOpen
  \bibfield  {author} {\bibinfo {author} {\bibnamefont {Iollo}, \bibfnamefont
  {A.}}, \bibinfo {author} {\bibnamefont {Dervieux}, \bibfnamefont {A.}},
  \bibinfo {author} {\bibnamefont {D{\'e}sid{\'e}ri}, \bibfnamefont {J.-A.}},
  and\ \bibinfo {author} {\bibnamefont {Lanteri}, \bibfnamefont {S.}},\
  }\bibfield  {title} {\enquote {\bibinfo {title} {Two stable pod-based
  approximations to the {N}avier--{S}tokes equations},}\ }\href@noop {}
  {\bibfield  {journal} {\bibinfo  {journal} {Computing and visualization in
  science}\ }\textbf {\bibinfo {volume} {3}},\ \bibinfo {pages} {61--66}
  (\bibinfo {year} {2000})}\BibitemShut {NoStop}%
\bibitem [{\citenamefont {Iwamoto}, \citenamefont {Suzuki},\ and\ \citenamefont
  {Kasagi}(2002{\natexlab{a}})}]{iwamoto2002database}%
  \BibitemOpen
  \bibfield  {author} {\bibinfo {author} {\bibnamefont {Iwamoto}, \bibfnamefont
  {K.}}, \bibinfo {author} {\bibnamefont {Suzuki}, \bibfnamefont {Y.}}, and\
  \bibinfo {author} {\bibnamefont {Kasagi}, \bibfnamefont {N.}},\ }\bibfield
  {title} {\enquote {\bibinfo {title} {Database of fully developed channel
  flow},}\ }\href@noop {} {\bibfield  {journal} {\bibinfo  {journal}
  {Department of Mechanical Engineering, The University of Tokyo, THTLAB
  Internal Report No. ILR-0201}\ } (\bibinfo {year}
  {2002}{\natexlab{a}})}\BibitemShut {NoStop}%
\bibitem [{\citenamefont {Iwamoto}, \citenamefont {Suzuki},\ and\ \citenamefont
  {Kasagi}(2002{\natexlab{b}})}]{iwamoto2002reynolds}%
  \BibitemOpen
  \bibfield  {author} {\bibinfo {author} {\bibnamefont {Iwamoto}, \bibfnamefont
  {K.}}, \bibinfo {author} {\bibnamefont {Suzuki}, \bibfnamefont {Y.}}, and\
  \bibinfo {author} {\bibnamefont {Kasagi}, \bibfnamefont {N.}},\ }\bibfield
  {title} {\enquote {\bibinfo {title} {Reynolds number effect on wall
  turbulence: toward effective feedback control},}\ }\href@noop {} {\bibfield
  {journal} {\bibinfo  {journal} {International journal of heat and fluid
  flow}\ }\textbf {\bibinfo {volume} {23}},\ \bibinfo {pages} {678--689}
  (\bibinfo {year} {2002}{\natexlab{b}})}\BibitemShut {NoStop}%
\bibitem [{\citenamefont {Kobayashi}\ and\ \citenamefont
  {Pereira}(1991)}]{Kobayashi1991}%
  \BibitemOpen
  \bibfield  {author} {\bibinfo {author} {\bibnamefont {Kobayashi},
  \bibfnamefont {M.~H.}}and\ \bibinfo {author} {\bibnamefont {Pereira},
  \bibfnamefont {J.~C.~F.}},\ }\bibfield  {title} {\enquote {\bibinfo {title}
  {Numerical comparison of momentum interpolation methods and
  pressure—velocity algorithms using non-staggered grids},}\ }\href
  {https://doi.org/https://doi.org/10.1002/cnm.1630070302} {\bibfield
  {journal} {\bibinfo  {journal} {Communications in Applied Numerical Methods}\
  }\textbf {\bibinfo {volume} {7}},\ \bibinfo {pages} {173--186} (\bibinfo
  {year} {1991})},\ \Eprint
  {https://arxiv.org/abs/https://onlinelibrary.wiley.com/doi/pdf/10.1002/cnm.1630070302}
  {https://onlinelibrary.wiley.com/doi/pdf/10.1002/cnm.1630070302} \BibitemShut
  {NoStop}%
\bibitem [{\citenamefont {Kostas}, \citenamefont {Soria},\ and\ \citenamefont
  {Chong}(2005)}]{kostas2005comparison}%
  \BibitemOpen
  \bibfield  {author} {\bibinfo {author} {\bibnamefont {Kostas}, \bibfnamefont
  {J.}}, \bibinfo {author} {\bibnamefont {Soria}, \bibfnamefont {J.}}, and\
  \bibinfo {author} {\bibnamefont {Chong}, \bibfnamefont {M.~S.}},\ }\bibfield
  {title} {\enquote {\bibinfo {title} {A comparison between snapshot {POD}
  analysis of {PIV} velocity and vorticity data},}\ }\href@noop {} {\bibfield
  {journal} {\bibinfo  {journal} {Experiments in fluids}\ }\textbf {\bibinfo
  {volume} {38}},\ \bibinfo {pages} {146--160} (\bibinfo {year}
  {2005})}\BibitemShut {NoStop}%
\bibitem [{\citenamefont {Lee}\ and\ \citenamefont {Dowell}(2020)}]{Lee2020}%
  \BibitemOpen
  \bibfield  {author} {\bibinfo {author} {\bibnamefont {Lee}, \bibfnamefont
  {M.~W.}}and\ \bibinfo {author} {\bibnamefont {Dowell}, \bibfnamefont
  {E.~H.}},\ }\bibfield  {title} {\enquote {\bibinfo {title} {{Improving the
  predictable accuracy of fluid {G}alerkin reduced-order models using two POD
  bases}},}\ }\href {https://doi.org/10.1007/s11071-020-05833-x} {\bibfield
  {journal} {\bibinfo  {journal} {Nonlinear Dynamics}\ }\textbf {\bibinfo
  {volume} {101}},\ \bibinfo {pages} {1457--1471} (\bibinfo {year}
  {2020})}\BibitemShut {NoStop}%
\bibitem [{\citenamefont {Lumley}(1967)}]{Lumley1967}%
  \BibitemOpen
  \bibfield  {author} {\bibinfo {author} {\bibnamefont {Lumley}, \bibfnamefont
  {J.~L.}},\ }\bibfield  {title} {\enquote {\bibinfo {title} {{The structure of
  inhomogeneous turbulent flows}},}\ }in\ \href@noop {} {\emph {\bibinfo
  {booktitle} {Proceedings of the international colloquium}}}\ (\bibinfo
  {publisher} {Publishing house Nakua},\ \bibinfo {address} {Moscow, USSR},\
  \bibinfo {year} {1967})\ pp.\ \bibinfo {pages} {166--167}\BibitemShut
  {NoStop}%
\bibitem [{\citenamefont {Michelsen}(1992)}]{michelsen1992basis3d}%
  \BibitemOpen
  \bibfield  {author} {\bibinfo {author} {\bibnamefont {Michelsen},
  \bibfnamefont {J.~A.}},\ }\bibfield  {title} {\enquote {\bibinfo {title}
  {Basis3{D} - a platform for development of multiblock {PDE} solvers},}\
  }\href@noop {} {\bibfield  {journal} {\bibinfo  {journal} {Report AFM}\
  }\textbf {\bibinfo {volume} {92}},\ \bibinfo {pages} {5} (\bibinfo {year}
  {1992})}\BibitemShut {NoStop}%
\bibitem [{\citenamefont {Michelsen}(1994)}]{michelsen1994block}%
  \BibitemOpen
  \bibfield  {author} {\bibinfo {author} {\bibnamefont {Michelsen},
  \bibfnamefont {J.~A.}},\ }\href@noop {} {\enquote {\bibinfo {title} {Block
  structured multigrid solution of 2{D} and 3{D} elliptic {PDE}'s},}\ }\bibinfo
  {type} {Tech. Rep.}\ (\bibinfo  {institution} {Dept.\ of Fluid Mechanics,
  Technical University of Denmark},\ \bibinfo {year} {1994})\BibitemShut
  {NoStop}%
\bibitem [{\citenamefont {Munir}\ \emph {et~al.}(2022)\citenamefont {Munir},
  \citenamefont {Siddiqui}, \citenamefont {bin Abdul~Aziz}, \citenamefont
  {Heikal},\ and\ \citenamefont {Farooq}}]{munir2022proper}%
  \BibitemOpen
  \bibfield  {author} {\bibinfo {author} {\bibnamefont {Munir}, \bibfnamefont
  {S.}}, \bibinfo {author} {\bibnamefont {Siddiqui}, \bibfnamefont {M.~I.}},
  \bibinfo {author} {\bibnamefont {bin Abdul~Aziz}, \bibfnamefont {A.~R.}},
  \bibinfo {author} {\bibnamefont {Heikal}, \bibfnamefont {M.}}, and\ \bibinfo
  {author} {\bibnamefont {Farooq}, \bibfnamefont {U.}},\ }\bibfield  {title}
  {\enquote {\bibinfo {title} {Proper orthogonal decomposition based on
  vorticity: Application in a two-phase slug flow},}\ }\href@noop {} {\bibfield
   {journal} {\bibinfo  {journal} {Journal of Fluids Engineering}\ }\textbf
  {\bibinfo {volume} {144}},\ \bibinfo {pages} {041501} (\bibinfo {year}
  {2022})}\BibitemShut {NoStop}%
\bibitem [{\citenamefont {Picard}\ and\ \citenamefont
  {Delville}(2000)}]{picard2000pressure}%
  \BibitemOpen
  \bibfield  {author} {\bibinfo {author} {\bibnamefont {Picard}, \bibfnamefont
  {C.}}and\ \bibinfo {author} {\bibnamefont {Delville}, \bibfnamefont {J.}},\
  }\bibfield  {title} {\enquote {\bibinfo {title} {Pressure velocity coupling
  in a subsonic round jet},}\ }\href@noop {} {\bibfield  {journal} {\bibinfo
  {journal} {International Journal of Heat and Fluid Flow}\ }\textbf {\bibinfo
  {volume} {21}},\ \bibinfo {pages} {359--364} (\bibinfo {year}
  {2000})}\BibitemShut {NoStop}%
\bibitem [{\citenamefont {Pope}(2000)}]{pope2000turbulent}%
  \BibitemOpen
  \bibfield  {author} {\bibinfo {author} {\bibnamefont {Pope}, \bibfnamefont
  {S.}},\ }\href@noop {} {\emph {\bibinfo {title} {Turbulent flows}}}\
  (\bibinfo  {publisher} {Cambridge university press},\ \bibinfo {year}
  {2000})\BibitemShut {NoStop}%
\bibitem [{\citenamefont {Richardson}(1922)}]{richardson1922weather}%
  \BibitemOpen
  \bibfield  {author} {\bibinfo {author} {\bibnamefont {Richardson},
  \bibfnamefont {L.~F.}},\ }\href@noop {} {\emph {\bibinfo {title} {Weather
  prediction by numerical process}}}\ (\bibinfo  {publisher} {University
  Press},\ \bibinfo {year} {1922})\BibitemShut {NoStop}%
\bibitem [{\citenamefont {Roman}\ \emph {et~al.}(2022)\citenamefont {Roman},
  \citenamefont {Campos}, \citenamefont {Romero},\ and\ \citenamefont
  {Tom{\'a}s}}]{roman2022slepc}%
  \BibitemOpen
  \bibfield  {author} {\bibinfo {author} {\bibnamefont {Roman}, \bibfnamefont
  {J.~E.}}, \bibinfo {author} {\bibnamefont {Campos}, \bibfnamefont {C.}},
  \bibinfo {author} {\bibnamefont {Romero}, \bibfnamefont {E.}}, and\ \bibinfo
  {author} {\bibnamefont {Tom{\'a}s}, \bibfnamefont {A.}},\ }\bibfield  {title}
  {\enquote {\bibinfo {title} {{SLEP}c users manual},}\ }\href@noop {}
  {\bibfield  {journal} {\bibinfo  {journal} {D. Sistemes Inform{\`a}tics i
  Computaci{\'o} Universitat Polit{\`e}cnica de Val{\`e}ncia, Valencia, Spain,
  Report No. DSIC-II/24/02}\ } (\bibinfo {year} {2022})}\BibitemShut {NoStop}%
\bibitem [{\citenamefont {Sengupta}(2012)}]{sengupta2012instabilities}%
  \BibitemOpen
  \bibfield  {author} {\bibinfo {author} {\bibnamefont {Sengupta},
  \bibfnamefont {T.~K.}},\ }\href@noop {} {\emph {\bibinfo {title}
  {Instabilities of flows and transition to turbulence}}}\ (\bibinfo
  {publisher} {CRC Press},\ \bibinfo {year} {2012})\BibitemShut {NoStop}%
\bibitem [{\citenamefont {Sengupta}\ \emph {et~al.}(2015)\citenamefont
  {Sengupta}, \citenamefont {Haider}, \citenamefont {Parvathi},\ and\
  \citenamefont {Pallavi}}]{sengupta2015enstrophy}%
  \BibitemOpen
  \bibfield  {author} {\bibinfo {author} {\bibnamefont {Sengupta},
  \bibfnamefont {T.~K.}}, \bibinfo {author} {\bibnamefont {Haider},
  \bibfnamefont {S.~I.}}, \bibinfo {author} {\bibnamefont {Parvathi},
  \bibfnamefont {M.~K.}}, and\ \bibinfo {author} {\bibnamefont {Pallavi},
  \bibfnamefont {G.}},\ }\bibfield  {title} {\enquote {\bibinfo {title}
  {Enstrophy-based proper orthogonal decomposition for reduced-order modeling
  of flow past a cylinder},}\ }\href@noop {} {\bibfield  {journal} {\bibinfo
  {journal} {Physical Review E}\ }\textbf {\bibinfo {volume} {91}},\ \bibinfo
  {pages} {043303} (\bibinfo {year} {2015})}\BibitemShut {NoStop}%
\bibitem [{\citenamefont {Shen}\ \emph {et~al.}(2003)\citenamefont {Shen},
  \citenamefont {Michelsen}, \citenamefont {Sørensen},\ and\ \citenamefont
  {Sørensen}}]{Shen2003}%
  \BibitemOpen
  \bibfield  {author} {\bibinfo {author} {\bibnamefont {Shen}, \bibfnamefont
  {W.~Z.}}, \bibinfo {author} {\bibnamefont {Michelsen}, \bibfnamefont
  {J.~A.}}, \bibinfo {author} {\bibnamefont {Sørensen}, \bibfnamefont
  {N.~N.}}, and\ \bibinfo {author} {\bibnamefont {Sørensen}, \bibfnamefont
  {J.~N.}},\ }\bibfield  {title} {\enquote {\bibinfo {title} {An improved
  simplec method on collocated grids for steady and unsteady flow
  computations},}\ }\href {https://doi.org/10.1080/713836202} {\bibfield
  {journal} {\bibinfo  {journal} {Numerical Heat Transfer, Part B:
  Fundamentals}\ }\textbf {\bibinfo {volume} {43}} (\bibinfo {year} {2003}),\
  10.1080/713836202}\BibitemShut {NoStop}%
\bibitem [{\citenamefont {Sirovich}(1987)}]{sirovich1987turbulence1}%
  \BibitemOpen
  \bibfield  {author} {\bibinfo {author} {\bibnamefont {Sirovich},
  \bibfnamefont {L.}},\ }\bibfield  {title} {\enquote {\bibinfo {title}
  {Turbulence and the dynamics of coherent structures. {I}. coherent
  structures},}\ }\href@noop {} {\bibfield  {journal} {\bibinfo  {journal}
  {Quarterly of applied mathematics}\ }\textbf {\bibinfo {volume} {45}},\
  \bibinfo {pages} {561--571} (\bibinfo {year} {1987})}\BibitemShut {NoStop}%
\bibitem [{\citenamefont {Sørensen}(1995)}]{sorensen1995}%
  \BibitemOpen
  \bibfield  {author} {\bibinfo {author} {\bibnamefont {Sørensen},
  \bibfnamefont {N.~N.}},\ }\emph {\bibinfo {title} {General purpose flow
  solver applied to flow over hills}},\ \href@noop {} {\bibinfo {type} {{PhD}
  thesis}},\ \bibinfo  {school} {Dept.\ of Meterology and Wind energy,
  Technical University of Denmark, and Risø National Laboratory} (\bibinfo
  {year} {1995})\BibitemShut {NoStop}%
\bibitem [{\citenamefont {Tardu}(2017)}]{tardu2017near}%
  \BibitemOpen
  \bibfield  {author} {\bibinfo {author} {\bibnamefont {Tardu}, \bibfnamefont
  {S.}},\ }\bibfield  {title} {\enquote {\bibinfo {title} {Near wall
  dissipation revisited},}\ }\href@noop {} {\bibfield  {journal} {\bibinfo
  {journal} {International Journal of Heat and Fluid Flow}\ }\textbf {\bibinfo
  {volume} {67}},\ \bibinfo {pages} {104--115} (\bibinfo {year}
  {2017})}\BibitemShut {NoStop}%
\bibitem [{\citenamefont {Ukeiley}\ \emph {et~al.}(1992)\citenamefont
  {Ukeiley}, \citenamefont {Varghese}, \citenamefont {Glauser},\ and\
  \citenamefont {Valentine}}]{ukeiley1992multifractal}%
  \BibitemOpen
  \bibfield  {author} {\bibinfo {author} {\bibnamefont {Ukeiley}, \bibfnamefont
  {L.}}, \bibinfo {author} {\bibnamefont {Varghese}, \bibfnamefont {M.}},
  \bibinfo {author} {\bibnamefont {Glauser}, \bibfnamefont {M.}}, and\ \bibinfo
  {author} {\bibnamefont {Valentine}, \bibfnamefont {D.}},\ }\bibfield  {title}
  {\enquote {\bibinfo {title} {Multifractal analysis of a lobed mixer flowfield
  utilizing the proper orthogonal decomposition},}\ }\href@noop {} {\bibfield
  {journal} {\bibinfo  {journal} {AIAA journal}\ }\textbf {\bibinfo {volume}
  {30}},\ \bibinfo {pages} {1260--1267} (\bibinfo {year} {1992})}\BibitemShut
  {NoStop}%
\bibitem [{\citenamefont {Weiss}(2019)}]{weiss2019tutorial}%
  \BibitemOpen
  \bibfield  {author} {\bibinfo {author} {\bibnamefont {Weiss}, \bibfnamefont
  {J.}},\ }\bibfield  {title} {\enquote {\bibinfo {title} {A tutorial on the
  proper orthogonal decomposition},}\ }in\ \href@noop {} {\emph {\bibinfo
  {booktitle} {AIAA Aviation 2019 Forum}}}\ (\bibinfo {year} {2019})\ p.\
  \bibinfo {pages} {3333}\BibitemShut {NoStop}%
\bibitem [{\citenamefont {Yang}\ \emph {et~al.}(2021)\citenamefont {Yang},
  \citenamefont {Hong}, \citenamefont {Lee},\ and\ \citenamefont
  {Huang}}]{YangDNS2021}%
  \BibitemOpen
  \bibfield  {author} {\bibinfo {author} {\bibnamefont {Yang}, \bibfnamefont
  {X.~I.~A.}}, \bibinfo {author} {\bibnamefont {Hong}, \bibfnamefont {J.}},
  \bibinfo {author} {\bibnamefont {Lee}, \bibfnamefont {M.}}, and\ \bibinfo
  {author} {\bibnamefont {Huang}, \bibfnamefont {X.~L.~D.}},\ }\bibfield
  {title} {\enquote {\bibinfo {title} {Grid resolution requirement for
  resolving rare and high intensity wall-shear stress events in direct
  numerical simulations},}\ }\href
  {https://doi.org/10.1103/PhysRevFluids.6.054603} {\bibfield  {journal}
  {\bibinfo  {journal} {Phys. Rev. Fluids}\ }\textbf {\bibinfo {volume} {6}},\
  \bibinfo {pages} {054603} (\bibinfo {year} {2021})}\BibitemShut {NoStop}%
\bibitem [{\citenamefont {Yeung}, \citenamefont {Donzis},\ and\ \citenamefont
  {Sreenivasan}(2012)}]{yeung2012dissipation}%
  \BibitemOpen
  \bibfield  {author} {\bibinfo {author} {\bibnamefont {Yeung}, \bibfnamefont
  {P.~K.}}, \bibinfo {author} {\bibnamefont {Donzis}, \bibfnamefont {D.~A.}},
  and\ \bibinfo {author} {\bibnamefont {Sreenivasan}, \bibfnamefont {K.~R.}},\
  }\bibfield  {title} {\enquote {\bibinfo {title} {Dissipation, enstrophy and
  pressure statistics in turbulence simulations at high {R}eynolds numbers},}\
  }\href@noop {} {\bibfield  {journal} {\bibinfo  {journal} {Journal of Fluid
  Mechanics}\ }\textbf {\bibinfo {volume} {700}},\ \bibinfo {pages} {5--15}
  (\bibinfo {year} {2012})}\BibitemShut {NoStop}%
\end{thebibliography}%


%aipnauth4-2.bst 2018-12-27 (MD) hand-edited version of apsauth4-1.bst
%Control: author (9) reversed initials

\end{document}